# A tentative view on hadrons

## V. A. Golovko

Moscow Polytechnic University

Bolshaya Semenovskaya 38, Moscow 107023, Russia

E-mail: fizika.mgvmi@mail.ru

Abstract

The present paper is based upon ideas and results obtained in an earlier paper by the author devoted to a new formulation of quantum electrodynamics. The equations derived in that paper depict the structure and properties of the electron as well as of other leptons including neutrinos. Since in nature there are only two stable charged elementary particles, the electron and proton (with their antiparticles), it is logical to suppose that in nature there is a protonic field too whose description is analogous with the one of the electronic field. The equations obtained in the paper and describing the proton and other baryons differ in some detail from the ones for the electronic field. A section of the paper is devoted to nuclei where the electronic and protonic fields are present simultaneously without neutrons. The neutron itself consists of an electron and a proton. Mesons are short-lived combinations of the electronic and protonic fields. The theory contains three dimensionless constants whose value can be calculated.

# 1. Introduction

Nowadays the hadrons are described by quantum chromodynamics (QCD). At the same time, QCD cannot be regarded as a genuine theory of elementary particles because it contains a number of arbitrary parameters, whereas the genuine theory should explain numerical values of all the relevant parameters. QCD is rather a phenomenological theory, the theory that postulates something or takes something from experiment and enables one to calculate something else. Besides, there are difficulties and many unresolved questions in QCD. A characteristic difficulty of QCD is the so-called proton spin puzzle [1]. In the initial quark model, the proton spin was built up from the spins of three quarks. Experiment shows, however, that the quarks carry only a small fraction ($\approx 1/3$) of the proton spin, and there are other contributions to it. It would be a miracle if the different irrational contributions yielded exactly ½ for the proton spin. The proton spin puzzle undermines the starting idea that the proton is composed of three quarks inasmuch as the idea is based first of all upon consideration of the spins of the proton and quarks.

In nature there are only two stable charged elementary particles, the electron and proton (with their antiparticles). In Ref. [2] a new formulation of quantum electrodynamics (QED) was proposed in which the electronic and electromagnetic fields are ordinary *c*-numbers in contradistinction to noncommuting *q*-numbers used in the standard formulation of QED (see also Ref. [3]). The electronic wave function (represented by two *c*-number bispinors in that case) describes the structure of the electron, which is not a point-like particle, and its motion in external fields. The same electronic field depicts also other leptons including neutrinos [2], which amounts to saying that the leptons are different manifestations of the electronic field. By analogy it is logical to suppose that in nature there is a protonic field as well whose description is similar to the one of the electronic field. At the same time equations for the protonic field must differ somehow from those of the electronic field because the electron magnetic moment is approximately equal to the Bohr magneton whereas the proton magnetic moment differs substantially from the nuclear magneton.

Seeing that there are no stable charged particles in nature apart from the electron and proton, all other elementary particles are to be described in terms of the electronic and protonic fields. Figuratively speaking, they should be composed of electrons and protons. For example, the neutron may consist of one electron and one proton. The principal objection to this is the spin: the electron and proton each have a spin equal to $\hbar/2$ so that they could not give $\hbar/2$ for the spin of the neutron. The proton, however, can so deform the electronic cloud that the spin of the cloud will no longer be equal to $\hbar/2$. An example of this is furnished by Eq. (4.3) of [3]: if we put $N = 1$, the spin of one electron may be different from $\hbar/2$ depending on the number ν. Another example follows from QCD: the proton spin is not equal to the sum of the spins of its three quarks (see above). Consequently, nuclei can be composed of electrons and protons. Historically, this was the first hypothesis concerning the structure of the nuclei, the hypothesis rejected because of a naïve consideration of the spins.

In Sec. 2 of the present paper, we derive and investigate equations that describe the protonic field leaning upon the methods and results of Ref. [2]. Section 3 is devoted to simultaneous description of the electronic and protonic fields. The results are applied to nuclei. In particular, we calculate the spin of the nuclei and compare it with experiment. In Sec. 4 we turn our attention to mesons. The results obtained are discussed in the concluding section.

Throughout the present paper, if the opposite is not pointed out, we utilize the definitions and notation adopted in [2] that coincide with the ones of Ref. [4]. It should be underlined that all quantities and functions in this paper are $c$-numbers.

## 2. Proton

As mentioned in Introduction a characteristic feature of the proton is the fact that its magnetic moment differs substantially from the nuclear magneton whereas the electron magnetic moment is approximately equal to the Bohr magneton. Pauli [5, Sec. 21] writes that the Dirac equation can be modified by adding a term of the type $F_{\mu\nu}\gamma^{\mu}\gamma^{\nu}\psi$. Seeing that the electromagnetic field tensor $F_{\mu\nu} = \partial A_{\nu}/\partial x^{\mu} - \partial A_{\mu}/\partial x^{\nu}$ is antisymmetric, one can take $F_{\mu\nu}\sigma^{\mu\nu}\psi$ instead of $F_{\mu\nu}\gamma^{\mu}\gamma^{\nu}\psi$ where

$$\sigma^{\mu\nu} = \frac{1}{2}(\gamma^{\mu}\gamma^{\nu} - \gamma^{\nu}\gamma^{\mu}). \qquad (2.1)$$

The Pauli term will yield an additional magnetic moment.

The electron charge will be denoted as $e$ with the sign ($e = -|e|$) and thereby the proton charge will be $-e$. The electronic field will be described by the bispinors $\psi_1$ and $\psi_2$ as in [2] while the protonic field by the bispinors $\Psi_1$ and $\Psi_2$. We imply the following gauge transformation

$$\psi_{1,2} \to \psi_{1,2}\exp\left(-\frac{ie}{c\hbar}f\right), \quad \Psi_{1,2} \to \Psi_{1,2}\exp\left(\frac{ie}{c\hbar}f\right), \quad A_{\mu} \to A_{\mu} + \frac{\partial f}{\partial x^{\mu}}. \qquad (2.2)$$

It was demonstrated in [2] that one should employ the Compton wavelength instead of the mass. When considering the proton separately from other particles it is natural to utilize the proton Compton wavelength $\lambdabar_{\mathrm{p}}$ as a standard of length replacing $m_{\mathrm{p}}$ by $\hbar/\lambdabar_{\mathrm{p}}c$. All of this enables one to write down the Lagrangian for the proton by analogy with Eq. (3.3) of [2]:

$$L(x) = -\frac{1}{16\pi} F_{\mu\nu} F^{\mu\nu} + \frac{ic\hbar}{2} \left( \overline{\Psi}_1 \gamma^\mu \frac{\partial \Psi_1}{\partial x^\mu} - \frac{\partial \overline{\Psi}_1}{\partial x^\mu} \gamma^\mu \Psi_1 - \overline{\Psi}_2 \gamma^\mu \frac{\partial \Psi_2}{\partial x^\mu} + \frac{\partial \overline{\Psi}_2}{\partial x^\mu} \gamma^\mu \Psi_2 \right)$$

$$+ eA_\mu (\overline{\Psi}_1 \gamma^\mu \Psi_1 - \overline{\Psi}_2 \gamma^\mu \Psi_2) + \frac{i}{2} e\chi F_{\mu\nu} (\overline{\Psi}_1 \sigma^{\mu\nu} \Psi_1 - \overline{\Psi}_2 \sigma^{\mu\nu} \Psi_2)$$

$$- \frac{c\hbar}{\lambda_p} (\overline{\Psi}_1 \Psi_1 - \overline{\Psi}_2 \Psi_2) + eV_\mu (\overline{\Psi}_1 \gamma^\mu \Psi_2 + \overline{\Psi}_2 \gamma^\mu \Psi_1) \,. \tag{2.3}$$

The signs here are chosen in compliance with the gauge transformation of (2.2). It is worthwhile to underline an essential distinction between the Lagrangian of (2.3) and the one of (3.3) of Ref. [2]: the former contains the four-potential $A_\mu$ and its derivatives whereas the latter does not contain the derivatives of $A_\mu$. The derivatives of $A_\mu$ figure in $F_{\mu\nu}$ of (2.3). The summand in (2.3) with $F_{\mu\nu}$ is relevant to the above Pauli term. The coefficient in the summand with a new constant $\chi$ is written in a form convenient for what follows.

Varying $L(x)$ with respect to $\overline{\Psi}_1$, $\overline{\Psi}_2$, $\Psi_1$ and $\Psi_2$ we obtain the following equations

$$ic\hbar\gamma^\mu \frac{\partial \Psi_1}{\partial x^\mu} + eA_\mu \gamma^\mu \Psi_1 - \frac{c\hbar}{\lambda_p} \Psi_1 + \frac{i}{2} e\chi F_{\mu\nu} \sigma^{\mu\nu} \Psi_1 + eV_\mu \gamma^\mu \Psi_2 = 0 \,, \tag{2.4}$$

$$ic\hbar\gamma^\mu \frac{\partial \Psi_2}{\partial x^\mu} + eA_\mu \gamma^\mu \Psi_2 - \frac{c\hbar}{\lambda_p} \Psi_2 + \frac{i}{2} e\chi F_{\mu\nu} \sigma^{\mu\nu} \Psi_2 - eV_\mu \gamma^\mu \Psi_1 = 0 \,, \tag{2.5}$$

$$ic\hbar \frac{\partial \overline{\Psi}_1}{\partial x^\mu} \gamma^\mu - eA_\mu \overline{\Psi}_1 \gamma^\mu + \frac{c\hbar}{\lambda_p} \overline{\Psi}_1 - \frac{i}{2} e\chi F_{\mu\nu} \overline{\Psi}_1 \sigma^{\mu\nu} - eV_\mu \overline{\Psi}_2 \gamma^\mu = 0 \,, \tag{2.6}$$

$$ic\hbar \frac{\partial \overline{\Psi}_2}{\partial x^\mu} \gamma^\mu - eA_\mu \overline{\Psi}_2 \gamma^\mu + \frac{c\hbar}{\lambda_p} \overline{\Psi}_2 - \frac{i}{2} e\chi F_{\mu\nu} \overline{\Psi}_2 \sigma^{\mu\nu} + eV_\mu \overline{\Psi}_1 \gamma^\mu = 0 \,. \tag{2.7}$$

Variation of $A_\mu$ in $L(x)$ yields

$$\frac{\partial F^{\mu\nu}}{\partial x^\nu} = 4\pi e (\overline{\Psi}_1 \gamma^\mu \Psi_1 - \overline{\Psi}_2 \gamma^\mu \Psi_2) + 4\pi i e\chi \frac{\partial}{\partial x^\nu} \left( \overline{\Psi}_1 \sigma^{\mu\nu} \Psi_1 - \overline{\Psi}_2 \sigma^{\mu\nu} \Psi_2 \right), \tag{2.8}$$

wherefrom the current density is

$$j^\mu = \overline{\Psi}_1 \gamma^\mu \Psi_1 - \overline{\Psi}_2 \gamma^\mu \Psi_2 + i\chi \frac{\partial}{\partial x^\nu} \left( \overline{\Psi}_1 \sigma^{\mu\nu} \Psi_1 - \overline{\Psi}_2 \sigma^{\mu\nu} \Psi_2 \right). \tag{2.9}$$

If we differentiate (2.8) with respect to $x^\mu$ and take the antisymmetry of $F^{\mu\nu}$ and $\sigma^{\mu\nu}$ into account, we obtain

$$\frac{\partial}{\partial x^\mu} \left( \overline{\Psi}_1 \gamma^\mu \Psi_1 - \overline{\Psi}_2 \gamma^\mu \Psi_2 \right) = 0 \,. \tag{2.10}$$

It should be remarked that this equation can be deduced from (2.4)–(2.7) as well. We see that the Pauli term does not contribute to the equation of continuity $\partial j^\mu/\partial x^\mu = 0$. Integrating Eq. (2.10)

over all space with account taken of the fact that there are no particles at infinity ($\Psi_1 = \Psi_2 = 0$) we have

$$\int_{(\infty)} (\Psi_1^* \Psi_1 - \Psi_2^* \Psi_2) dV = I \, , \qquad (2.11)$$

where $I = $ constant. The normalization for the proton will be $I = 1$.

Finally, varying $L(x)$ with respect to $V_\mu$ one obtains

$$\overline{\Psi}_1 \gamma^\mu \Psi_2 + \overline{\Psi}_2 \gamma^\mu \Psi_1 = 0 \, . \qquad (2.12)$$

It is not difficult to deduce from Eqs. (2.4)–(2.7) that

$$\frac{\partial}{\partial x^\mu} \left( \overline{\Psi}_1 \gamma^\mu \Psi_2 + \overline{\Psi}_2 \gamma^\mu \Psi_1 \right) = 0 \, . \qquad (2.13)$$

Therefore the condition of (2.12) does not contradict the equations of motion of (2.4)–(2.7).

We now turn to the energy-momentum tensor $T^{\mu\nu}$. Proceeding as in Sec. 3 of [2] we obtain the tensor in a symmetric and gauge invariant form:

$$T^{\mu\nu} = \frac{1}{16\pi} \left( g^{\mu\nu} F_{\lambda\eta} F^{\lambda\eta} - 4 F^{\mu\lambda} F^\nu{}_\lambda \right) + \frac{e}{2} \left( A^\mu \overline{\Psi}_1 \gamma^\nu \Psi_1 + A^\nu \overline{\Psi}_1 \gamma^\mu \Psi_1 - A^\mu \overline{\Psi}_2 \gamma^\nu \Psi_2 - A^\nu \overline{\Psi}_2 \gamma^\mu \Psi_2 \right)$$

$$- \frac{ie\chi}{2} \left( F^\mu{}_\eta \overline{\Psi}_1 \sigma^{\eta\nu} \Psi_1 + F^\nu{}_\eta \overline{\Psi}_1 \sigma^{\eta\mu} \Psi_1 - F^\mu{}_\eta \overline{\Psi}_2 \sigma^{\eta\nu} \Psi_2 - F^\nu{}_\eta \overline{\Psi}_2 \sigma^{\eta\mu} \Psi_2 \right)$$

$$+ \frac{ic\hbar}{4} \left( \overline{\Psi}_1 \gamma^\nu \frac{\partial \Psi_1}{\partial x_\mu} + \overline{\Psi}_1 \gamma^\mu \frac{\partial \Psi_1}{\partial x_\nu} - \frac{\partial \overline{\Psi}_1}{\partial x_\mu} \gamma^\nu \Psi_1 - \frac{\partial \overline{\Psi}_1}{\partial x_\nu} \gamma^\mu \Psi_1 - \overline{\Psi}_2 \gamma^\nu \frac{\partial \Psi_2}{\partial x_\mu} - \overline{\Psi}_2 \gamma^\mu \frac{\partial \Psi_2}{\partial x_\nu} \right.$$

$$\left. + \frac{\partial \overline{\Psi}_2}{\partial x_\mu} \gamma^\nu \Psi_2 + \frac{\partial \overline{\Psi}_2}{\partial x_\nu} \gamma^\mu \Psi_2 \right) . \qquad (2.14)$$

From this we can calculate the density of energy $T^{00}$:

$$T^{00} = \frac{E^2 + H^2}{8\pi} + e\varphi(\Psi_1^* \Psi_1 - \Psi_2^* \Psi_2) + e\chi \mathbf{E} \mathbf{\Gamma} + \frac{i\hbar}{2} \left( \Psi_1^* \frac{\partial \Psi_1}{\partial t} - \frac{\partial \Psi_1^*}{\partial t} \Psi_1 - \Psi_2^* \frac{\partial \Psi_2}{\partial t} + \frac{\partial \Psi_2^*}{\partial t} \Psi_2 \right) . \qquad (2.15)$$

Here we have used the three-dimensional notation as in Eq. (3.19) of [2] and

$$\mathbf{\Gamma} = i(\Psi_1^* \boldsymbol{\gamma} \Psi_1 - \Psi_2^* \boldsymbol{\gamma} \Psi_2) = i(\overline{\Psi}_1 \boldsymbol{\alpha} \Psi_1 - \overline{\Psi}_2 \boldsymbol{\alpha} \Psi_2) \, . \qquad (2.16)$$

By analogy with Eq. (3.21) of [2] we can compute the vector $\mathbf{M}$ of angular momentum

$$\mathbf{M} = \frac{1}{4\pi c} \int_{(\infty)} [\mathbf{r}[\mathbf{E}\mathbf{H}]] dV + \frac{e}{c} \int_{(\infty)} \left( [\mathbf{r}\mathbf{A}](\overline{\Psi}_1 \gamma^0 \Psi_1 - \overline{\Psi}_2 \gamma^0 \Psi_2) - \chi[\mathbf{r}[\mathbf{H}\mathbf{\Gamma}]] \right) dV$$

$$+ \hbar \int_{(\infty)} \left\{ \frac{1}{2} \left( \overline{\Psi}_1 \gamma^0 \boldsymbol{\Sigma} \Psi_1 - \overline{\Psi}_2 \gamma^0 \boldsymbol{\Sigma} \Psi_2 \right) - i \left( \overline{\Psi}_1 \gamma^0 [\mathbf{r}\nabla\Psi_1] - \overline{\Psi}_2 \gamma^0 [\mathbf{r}\nabla\Psi_2] \right) \right\} dV \, . \qquad (2.17)$$

The question as to how the equations should be modified if the system is in an external electromagnetic field represented by a four-potential $A_\mu^{\text{ext}}$ can be considered as in Sec. 3 of [2]. One must replace $A_\mu$ with $A_\mu + A_\mu^{\text{ext}}$ in Eqs. (2.4)–(2.7) while Eq. (2.8) remains unchanged. The energy of the system in the external field assumed to be time-independent, $\mathscr{E} = \int T^{00} dV$, can be calculated with use made of (2.15) without adding $A_\mu^{\text{ext}}$ hereinto.

Reverting to the current density four-vector $j^\mu$ of (2.9) it will be useful to write down separately the time component $j^0$ representing the number density of matter $\rho_{\text{m}}$ and the current density vector $\mathbf{j}$:

$$\rho_{\text{m}} = \Psi_1^* \Psi_1 - \Psi_2^* \Psi_2 + \chi \operatorname{div} \mathbf{\Gamma}, \quad \mathbf{j} = \Psi_1^* \boldsymbol{\alpha} \Psi_1 - \Psi_2^* \boldsymbol{\alpha} \Psi_2 - \chi \frac{\partial \mathbf{\Gamma}}{c \partial t} - \chi \operatorname{curl} \mathbf{\Delta}, \tag{2.18}$$

where

$$\mathbf{\Delta} = -\left( \overline{\Psi}_1 \mathbf{\Sigma} \Psi_1 - \overline{\Psi}_2 \mathbf{\Sigma} \Psi_2 \right). \tag{2.19}$$

Integrating $\rho_{\text{m}}$ over all space one obtains (2.11).

### 2.1. Dimensionless equations

Since Eqs. (2.6)–(2.7) are complex conjugate with respect to Eqs. (2.4)–(2.5), it is sufficient to work with the latter. We introduce dimensionless quantities, denoting them by a tilde, according to

$$x^\mu = \lambdabar_{\text{p}} \tilde{x}^\mu, \ \Psi_{1,2} = \frac{1}{\lambdabar_{\text{p}}^{3/2}} \tilde{\Psi}_{1,2}, \ A_\mu = \frac{-e}{\lambdabar_{\text{p}}} \tilde{A}_\mu, \ V_\mu = \frac{-e}{\lambdabar_{\text{p}}} \tilde{V}_\mu, \ \kappa = \frac{\chi}{\lambdabar_{\text{p}}}; \quad \mathbf{E} = \frac{-e}{\lambdabar_{\text{p}}^2} \tilde{\mathbf{E}}, \ \mathbf{H} = \frac{-e}{\lambdabar_{\text{p}}^2} \tilde{\mathbf{H}}. \tag{2.20}$$

Note that the first equality at $\mu = 0$ reads $ct = \lambdabar_{\text{p}} \tilde{t}$. We recast Eqs. (2.4)–(2.5) in the three-dimensional notation taking into account that $A_\mu = (\varphi, -\mathbf{A})$ and $V_\mu = (V_0, -\mathbf{V})$:

$$i \frac{\partial \tilde{\Psi}_1}{\partial \tilde{t}} + i \boldsymbol{\alpha} \tilde{\nabla} \tilde{\Psi}_1 + \alpha \tilde{\mathbf{A}} \boldsymbol{\alpha} \tilde{\Psi}_1 - \alpha \tilde{\varphi} \tilde{\Psi}_1 - \beta \tilde{\Psi}_1 - i \alpha \kappa \left( \tilde{\mathbf{E}} \boldsymbol{\gamma} + \tilde{\mathbf{H}} \boldsymbol{\beta} \right) \tilde{\Psi}_1 - \alpha \tilde{V}_0 \tilde{\Psi}_2 + \alpha \tilde{\mathbf{V}} \boldsymbol{\alpha} \tilde{\Psi}_2 = 0, \tag{2.21}$$

$$i \frac{\partial \tilde{\Psi}_2}{\partial \tilde{t}} + i \boldsymbol{\alpha} \tilde{\nabla} \tilde{\Psi}_2 + \alpha \tilde{\mathbf{A}} \boldsymbol{\alpha} \tilde{\Psi}_2 - \alpha \tilde{\varphi} \tilde{\Psi}_2 - \beta \tilde{\Psi}_2 - i \alpha \kappa \left( \tilde{\mathbf{E}} \boldsymbol{\gamma} + \tilde{\mathbf{H}} \boldsymbol{\beta} \right) \tilde{\Psi}_2 + \alpha \tilde{V}_0 \tilde{\Psi}_1 - \alpha \tilde{\mathbf{V}} \boldsymbol{\alpha} \tilde{\Psi}_1 = 0. \tag{2.22}$$

Here $\alpha = e^2/\hbar c$ is the fine-structure constant and $\boldsymbol{\beta} = \frac{1}{2}[\boldsymbol{\gamma} \boldsymbol{\alpha}] = i \gamma^0 \mathbf{\Sigma}$ {cf. Eq. (21.22) of [4]}. Equation (2.8) yields two equations [cf. (2.18)– (2.19)]

$$\nabla^2 \tilde{\varphi} + \frac{\partial}{\partial \tilde{t}} \operatorname{div} \tilde{\mathbf{A}} + 4\pi \left( \tilde{\Psi}_1^* \tilde{\Psi}_1 - \tilde{\Psi}_2^* \tilde{\Psi}_2 \right) + 4\pi \kappa \operatorname{div} \tilde{\mathbf{\Gamma}} = 0, \tag{2.23}$$

$$\nabla^2 \tilde{\mathbf{A}} - \frac{\partial^2 \tilde{\mathbf{A}}}{\partial \tilde{t}^2} - \nabla \operatorname{div} \tilde{\mathbf{A}} - \nabla \frac{\partial \tilde{\varphi}}{\partial \tilde{t}} + 4\pi \left( \tilde{\Psi}_1^* \boldsymbol{\alpha} \tilde{\Psi}_1 - \tilde{\Psi}_2^* \boldsymbol{\alpha} \tilde{\Psi}_2 \right) - 4\pi \kappa \frac{\partial \tilde{\mathbf{\Gamma}}}{\partial \tilde{t}} - 4\pi \kappa \operatorname{curl} \tilde{\mathbf{\Delta}} = 0. \tag{2.24}$$

In what follows we shall omit the tilde bearing in mind that, if an equation does not contain $\hbar$, $c$, $e$, $\chi$ or $\lambda_p$ but contains $\alpha$ or $\kappa$, the equation is relevant to the dimensionless quantities.

We look for stationary solutions. As in [2] we neglect the factor $\exp(-i\varepsilon t/\hbar)$ that figures in the wave function in this instance upon assuming that the energy $\varepsilon$ is added to the unknown constant part of $\varphi$ [in effect, this is a gauge transformation according to (2.2)]. For this reason we merely put $\partial/\partial t = 0$ when dealing with the stationary solutions. We seek also axially symmetric solutions. Analogously to [2] the solution to Eqs. (2.21) and (2.22) in the cylindrical coordinates $\rho$, $\dot{\varphi}$, $z$ can be sought in the form (to distinguish between the angle $\varphi$ and the potential $\varphi$ we mark the angle with a dot at the top)

$$\Psi_1 = \begin{pmatrix} F_1^{(1)}(\rho,z) \\ F_2^{(1)}(\rho,z)e^{i\dot{\varphi}} \\ iF_3^{(1)}(\rho,z) \\ iF_4^{(1)}(\rho,z)e^{i\dot{\varphi}} \end{pmatrix}, \qquad \Psi_2 = \begin{pmatrix} F_1^{(2)}(\rho,z) \\ F_2^{(2)}(\rho,z)e^{i\dot{\varphi}} \\ iF_3^{(2)}(\rho,z) \\ iF_4^{(2)}(\rho,z)e^{i\dot{\varphi}} \end{pmatrix}, \qquad (2.25)$$

in which the functions $F_i^{(1,2)}$ are real-valued.

Now the vector $\boldsymbol{\Gamma}$ of (2.16) has the component $\Gamma_{\dot{\varphi}} = 0$ while two other components are

$$\Gamma_\rho = 2\left(F_1^{(2)}F_4^{(2)} + F_2^{(2)}F_3^{(2)} - F_1^{(1)}F_4^{(1)} - F_2^{(1)}F_3^{(1)}\right),$$
$$\Gamma_z = 2\left(F_2^{(1)}F_4^{(1)} - F_1^{(1)}F_3^{(1)} - F_2^{(2)}F_4^{(2)} + F_1^{(2)}F_3^{(2)}\right). \qquad (2.26)$$

The vector $\boldsymbol{\Delta}$ of (2.19) has also the component $\Delta_{\dot{\varphi}} = 0$ while its other components are

$$\Delta_\rho = 2\left(F_3^{(1)}F_4^{(1)} - F_1^{(1)}F_2^{(1)} - F_3^{(2)}F_4^{(2)} + F_1^{(2)}F_2^{(2)}\right),$$
$$\Delta_z = -F_1^{(1)^2} + F_2^{(1)^2} + F_3^{(1)^2} - F_4^{(1)^2} + F_1^{(2)^2} - F_2^{(2)^2} - F_3^{(2)^2} + F_4^{(2)^2}. \qquad (2.27)$$

Since these components do not depend on $\dot{\varphi}$, the curl of $\boldsymbol{\Delta}$ that figures in (2.18) has only the $\dot{\varphi}$-component equal to $\partial\Delta_\rho/\partial z - \partial\Delta_z/\partial\rho$. The first two terms in the formula for $\mathbf{j}$ of (2.18) have only the $\dot{\varphi}$-components as well. Therefore there is only one nonzero component of the current density equal to, in the dimensionless form,

$$j_{\dot{\varphi}} = 2\left(F_1^{(1)}F_4^{(1)} - F_2^{(1)}F_3^{(1)} - F_1^{(2)}F_4^{(2)} + F_2^{(2)}F_3^{(2)}\right) - \kappa\left(\frac{\partial\Delta_\rho}{\partial z} - \frac{\partial\Delta_z}{\partial\rho}\right). \qquad (2.28)$$

Seeing that the current density $\mathbf{j}$ figures in Eq. (2.24) for the vector potential $\mathbf{A}$, the vector $\mathbf{A}$ has only one nonzero component $A_{\dot{\varphi}}$ as well (denoted henceforth as $A$). This component does not depend on the angle $\dot{\varphi}$ and thereby div $\mathbf{A} = 0$ in the present situation. Along similar lines, we presume that the vector $\mathbf{V}$ has only one nonzero component $V_{\dot{\varphi}} = V$, too. The field intensities $\mathbf{E}$ and $\mathbf{H}$ ($\mathbf{E} = -\nabla\varphi$, $\mathbf{H} = \mathrm{curl}\,\mathbf{A}$) have the following components

$$E_\rho = -\frac{\partial \varphi}{\partial \rho}, \, E_z = -\frac{\partial \varphi}{\partial z}, \, E_\phi = 0; \quad H_\rho = -\frac{\partial A}{\partial z}, \, H_z = \frac{\partial A}{\partial \rho} + \frac{A}{\rho}, \, H_{\hat\phi} = 0 \,. \tag{2.29}$$

Now Eqs. (2.21) and (2.22) in the cylindrical coordinates take the form (we use the standard representation [4] for the matrices)

$$\frac{\partial F_4^{(1)}}{\partial \rho} + \frac{\partial F_3^{(1)}}{\partial z} + \frac{F_4^{(1)}}{\rho} + (1 + \alpha\varphi - \alpha\kappa H_z)F_1^{(1)} - \alpha\kappa H_\rho F_2^{(1)} - \alpha\kappa E_z F_3^{(1)} - \alpha(A + \kappa E_\rho)F_4^{(1)}$$
$$+ \alpha V_0 F_1^{(2)} - \alpha V F_4^{(2)} = 0, \tag{2.30}$$

$$\frac{\partial F_3^{(1)}}{\partial \rho} - \frac{\partial F_4^{(1)}}{\partial z} - \alpha\kappa H_\rho F_1^{(1)} + (1 + \alpha\varphi + \alpha\kappa H_z)F_2^{(1)} + \alpha(A - \kappa E_\rho)F_3^{(1)} + \alpha\kappa E_z F_4^{(1)}$$
$$+ \alpha V_0 F_2^{(2)} + \alpha V F_3^{(2)} = 0, \tag{2.31}$$

$$\frac{\partial F_2^{(1)}}{\partial \rho} + \frac{\partial F_1^{(1)}}{\partial z} + \frac{F_2^{(1)}}{\rho} + \alpha\kappa E_z F_1^{(1)} - \alpha(A - \kappa E_\rho)F_2^{(1)} + (1 - \alpha\varphi - \alpha\kappa H_z)F_3^{(1)} - \alpha\kappa H_\rho F_4^{(1)}$$
$$- \alpha V_0 F_3^{(2)} - \alpha V F_2^{(2)} = 0, \tag{2.32}$$

$$\frac{\partial F_1^{(1)}}{\partial \rho} - \frac{\partial F_2^{(1)}}{\partial z} + \alpha(A + \kappa E_\rho)F_1^{(1)} - \alpha\kappa E_z F_2^{(1)} - \alpha\kappa H_\rho F_3^{(1)} + (1 - \alpha\varphi + \alpha\kappa H_z)F_4^{(1)}$$
$$- \alpha V_0 F_4^{(2)} + \alpha V F_1^{(2)} = 0, \tag{2.33}$$

$$\frac{\partial F_4^{(2)}}{\partial \rho} + \frac{\partial F_3^{(2)}}{\partial z} + \frac{F_4^{(2)}}{\rho} + (1 + \alpha\varphi - \alpha\kappa H_z)F_1^{(2)} - \alpha\kappa H_\rho F_2^{(2)} - \alpha\kappa E_z F_3^{(2)} - \alpha(A + \kappa E_\rho)F_4^{(2)}$$
$$- \alpha V_0 F_1^{(1)} + \alpha V F_4^{(1)} = 0, \tag{2.34}$$

$$\frac{\partial F_3^{(2)}}{\partial \rho} - \frac{\partial F_4^{(2)}}{\partial z} - \alpha\kappa H_\rho F_1^{(2)} + (1 + \alpha\varphi + \alpha\kappa H_z)F_2^{(2)} + \alpha(A - \kappa E_\rho)F_3^{(2)} + \alpha\kappa E_z F_4^{(2)}$$
$$- \alpha V_0 F_2^{(1)} - \alpha V F_3^{(1)} = 0, \tag{2.35}$$

$$\frac{\partial F_2^{(2)}}{\partial \rho} + \frac{\partial F_1^{(2)}}{\partial z} + \frac{F_2^{(2)}}{\rho} + \alpha\kappa E_z F_1^{(2)} - \alpha(A - \kappa E_\rho)F_2^{(2)} + (1 - \alpha\varphi - \alpha\kappa H_z)F_3^{(2)} - \alpha\kappa H_\rho F_4^{(2)}$$
$$+ \alpha V_0 F_3^{(1)} + \alpha V F_2^{(1)} = 0, \tag{2.36}$$

$$\frac{\partial F_1^{(2)}}{\partial \rho} - \frac{\partial F_2^{(2)}}{\partial z} + \alpha(A + \kappa E_\rho)F_1^{(2)} - \alpha\kappa E_z F_2^{(2)} - \alpha\kappa H_\rho F_3^{(2)} + (1 - \alpha\varphi + \alpha\kappa H_z)F_4^{(2)}$$
$$+ \alpha V_0 F_4^{(1)} - \alpha V F_1^{(1)} = 0. \tag{2.37}$$

Equations (2.23) and (2.24) become

$$\nabla^2\varphi + 4\pi\left( F_1^{(1)^2} + F_2^{(1)^2} + F_3^{(1)^2} + F_4^{(1)^2} - F_1^{(2)^2} - F_2^{(2)^2} - F_3^{(2)^2} - F_4^{(2)^2} \right)$$

$$+ 4\pi\kappa\left(\frac{\partial\Gamma_\rho}{\partial\rho} + \frac{\Gamma_\rho}{\rho} + \frac{\partial\Gamma_z}{\partial z}\right) = 0\,, \tag{2.38}$$

$$\nabla^2 A - \frac{A}{\rho^2} + 8\pi\left(F_1^{(1)}F_4^{(1)} - F_2^{(1)}F_3^{(1)} - F_1^{(2)}F_4^{(2)} + F_2^{(2)}F_3^{(2)}\right) - 4\pi\kappa\left(\frac{\partial\Delta_\rho}{\partial z} - \frac{\partial\Delta_z}{\partial\rho}\right) = 0\,. \tag{2.39}$$

The conditions of (2.12) furnish two equations

$$F_1^{(1)}F_1^{(2)} + F_2^{(1)}F_2^{(2)} + F_3^{(1)}F_3^{(2)} + F_4^{(1)}F_4^{(2)} = 0\,, \tag{2.40}$$

$$F_1^{(1)}F_4^{(2)} - F_2^{(1)}F_3^{(2)} - F_3^{(1)}F_2^{(2)} + F_4^{(1)}F_1^{(2)} = 0\,. \tag{2.41}$$

The equations must be supplemented with the normalization condition that follows from (2.11):

$$\int\limits_{(\infty)}\left(F_1^{(1)^2} + F_2^{(1)^2} + F_3^{(1)^2} + F_4^{(1)^2} - F_1^{(2)^2} - F_2^{(2)^2} - F_3^{(2)^2} - F_4^{(2)^2}\right)dV = 1\,. \tag{2.42}$$

We discuss now properties of the formation described by the above equations. It will be shown below that the functions $F_i^{(1,2)}$ decrease exponentially with the distance from the center so that they correspond to a particle-like structure. It will be noted that we presume that none of the occurring functions have singularities anywhere the point $r = 0$ inclusive.

We rewrite Eq. (2.38) in spherical coordinates $r$, $\theta$, $\dot\phi$ analogously with Eq. (4.19) of [2]. When integrated over all space the term with the coefficient $\kappa$ vanishes because it is proportional to the divergence of $\Gamma$ according to (2.23). As a result we arrive at Eq. (4.20) of [2] from which and Eq. (2.42) it follows that $\phi \to 1/r$ + constant as $r \to \infty$, or $\phi \to -e/r$ + constant in the dimensional units according to (2.20). The last asymptotic expression indicates that the formation in question has the same charge $-e$ as the proton.

The magnetic moment $\mu$ in the dimensionless units is defined by Eq. (4.21) of [2]. Analogously with Eq. (4.22) of [2], with use made of (2.28) we find that $\mu$ has only one component

$$\mu_z = \int\limits_{(\infty)}\rho\left[F_1^{(1)}F_4^{(1)} - F_2^{(1)}F_3^{(1)} - F_1^{(2)}F_4^{(2)} + F_2^{(2)}F_3^{(2)} - \frac{\kappa}{2}\left(\frac{\partial\Delta_\rho}{\partial z} - \frac{\partial\Delta_z}{\partial\rho}\right)\right]dV\,. \tag{2.43}$$

To calculate $\mu_z$ it is necessary to solve the above equations. We shall revert to this question below in Sec. 2.2.

We turn now to the angular momentum of the formation. Similarly to [2] it can be shown that the first two integrals in (2.17) vanish for the stationary and axially symmetric solution under consideration if Eq. (2.23) is taken into account. Upon introducing (2.25) into the third integral and integrating over the angle $\dot\phi$ we can see that there is only one nonzero component of the vector $\mathbf{M}$ equal, in view of (2.42), to $M_z = \hbar/2$ in the ordinary units. Therefore the explanation of the proton spin $\hbar/2$ encounters no difficulties in the present approach in contradistinction to QCD (see Introduction).

In Ref. [2] a hypothesis is put forward according to which the equations obtained in [2] have a particle-like solution relevant to the electron only if the fine-structure constant $\alpha$ has the well-known experimental value ($\alpha \approx 1/137$). Seeing that the value of $\alpha$ is now predetermined, it is reasonable to hypothesize that the above equations have a particle-like and physically admissible (without singularities) solution, the solution relevant to the proton, only if the second dimensionless constant in the equations, $\kappa$, has a well-defined value. Of course the exact value of $\kappa$ and the resultant solution can be found only numerically. An approximate value of $\kappa$ will be calculated below in Sec. 2.2.

We now proceed to a further analysis of the equations obtained. As in [2] we assume the following structure of the functions involved

$$F_1^{(1)} = G_1^{(1)}(\rho^2, z^2), F_2^{(1)} = \rho z G_2^{(1)}(\rho^2, z^2), F_3^{(1)} = z G_3^{(1)}(\rho^2, z^2), F_4^{(1)} = \rho G_4^{(1)}(\rho^2, z^2),$$

$$F_1^{(2)} = z G_1^{(2)}(\rho^2, z^2), F_2^{(2)} = \rho G_2^{(2)}(\rho^2, z^2), F_3^{(2)} = G_3^{(2)}(\rho^2, z^2), F_4^{(2)} = \rho z G_4^{(2)}(\rho^2, z^2),$$

$$\varphi = \varphi(\rho^2, z^2), A = \rho \gamma(\rho^2, z^2), V_0 = z U_0(\rho^2, z^2), V = \rho z U(\rho^2, z^2). \qquad (2.44)$$

In addition we write down the structure of the field intensities **E** and **H** which is clear from (2.29):

$$E_\rho = \rho E_1(\rho^2, z^2), E_1 = -\frac{\partial \varphi}{\rho \partial \rho}, E_z = z E_2(\rho^2, z^2), E_2 = -\frac{\partial \varphi}{z \partial z},$$

$$H_\rho = \rho z H_1(\rho^2, z^2), H_1 = -\frac{\partial \gamma}{z \partial z}, H_z = H_2(\rho^2, z^2), H_2 = 2\gamma + \rho \frac{\partial \gamma}{\partial \rho}. \qquad (2.45)$$

It will be recalled that the energy $\varepsilon$ that figures in the factor $\exp(-i\varepsilon t/\hbar)$ for stationary states is added to the constant part of the potential $\varphi$. It is preferable to single out the relativistic rest energy $m_p c^2$ explicitly in this part. Proceeding as in [2] we shall obtain in parallel with (4.26) of [2] that

$$\tilde{\varphi} = -\frac{1}{\alpha} + C_0 + \overline{\varphi}(\mathbf{r}), \qquad (2.46)$$

where $C_0$ is an unknown constant which should be small as compared with $1/\alpha \approx 137$ and $\overline{\varphi}(\mathbf{r}) \to 0$ as $r \to \infty$.

To verify the rationality of (2.46) we can find the energy of the formation $\mathscr{E} = \int T^{00} dV$ with use made of (2.15) where $\partial/\partial t = 0$ and with account taken of (2.46) and (2.11) where $I = 1$ [see also (2.42)]. In the ordinary units

$$\mathcal{E} = \frac{c\hbar}{\lambdabar_p}\left\{1 - \alpha C_0 + \alpha \int\limits_{(\infty)}\left[\frac{1}{8\pi}\left(\tilde{E}^2 + \tilde{H}^2\right) - \overline{\varphi}\left(\tilde{\Psi}_1^*\tilde{\Psi}_1 - \tilde{\Psi}_2^*\tilde{\Psi}_2\right) - \kappa\tilde{\mathbf{E}}\tilde{\boldsymbol{\Gamma}}\right]d\tilde{V}\right\}. \tag{2.47}$$

For clarity sake, here again we have utilized the tilde to denote the dimensionless quantities of (2.20). Putting $\alpha = 0$ we obtain that $\mathcal{E} = m_p c^2$ if $\lambdabar_p = \hbar/m_p c$ as is usual.

We can now write down the equations for the new functions that follow from (2.30)–(2.37):

$$\rho\frac{\partial G_4^{(1)}}{\partial\rho} + z\frac{\partial G_3^{(1)}}{\partial z} + \left[\alpha\left(C_0 + \overline{\varphi}\right) - \alpha\kappa H_2\right]G_1^{(1)} - \alpha\kappa\rho^2 z^2 H_1 G_2^{(1)} + \left(1 - \alpha\kappa z^2 E_2\right)G_3^{(1)}$$
$$+ \left[2 - \alpha\rho^2\left(\gamma + \kappa E_1\right)\right]G_4^{(1)} + \alpha z^2 U_0 G_1^{(2)} - \alpha\rho^2 z^2 U G_4^{(2)} = 0, \tag{2.48}$$

$$\frac{\partial G_3^{(1)}}{\rho\partial\rho} - \frac{\partial G_4^{(1)}}{z\partial z} - \alpha\kappa H_1 G_1^{(1)} + \left[\alpha\left(C_0 + \overline{\varphi}\right) + \alpha\kappa H_2\right]G_2^{(1)} + \alpha\left(\gamma - \kappa E_1\right)G_3^{(1)} + \alpha\kappa E_2\, G_4^{(1)}$$
$$+ \alpha U_0 G_2^{(2)} + \alpha U G_3^{(2)} = 0, \tag{2.49}$$

$$\rho\frac{\partial G_2^{(1)}}{\partial\rho} + \frac{\partial G_1^{(1)}}{z\partial z} + \alpha\kappa E_2\, G_1^{(1)} + \left[2 - \alpha\rho^2\left(\gamma - \kappa E_1\right)\right]G_2^{(1)} + \left(2 - \alpha C_0 - \alpha\overline{\varphi} - \alpha\kappa H_2\right)G_3^{(1)}$$
$$- \alpha\kappa\rho^2 H_1\, G_4^{(1)} - \alpha U_0 G_3^{(2)} - \alpha\rho^2 U G_2^{(2)} = 0, \tag{2.50}$$

$$\frac{\partial G_1^{(1)}}{\rho\partial\rho} - z\frac{\partial G_2^{(1)}}{\partial z} + \alpha\left(\gamma + \kappa E_1\right)G_1^{(1)} - \left(1 + \alpha\kappa z^2 E_2\right)G_2^{(1)} - \alpha\kappa z^2 H_1 G_3^{(1)} + \left(2 - \alpha C_0 - \alpha\overline{\varphi} + \alpha\kappa H_2\right)G_4^{(1)}$$
$$- \alpha z^2 U_0 G_4^{(2)} + \alpha z^2 U G_1^{(2)} = 0, \tag{2.51}$$

$$\rho\frac{\partial G_4^{(2)}}{\partial\rho} + \frac{\partial G_3^{(2)}}{z\partial z} + \left[\alpha\left(C_0 + \overline{\varphi}\right) - \alpha\kappa H_2\right]G_1^{(2)} - \alpha\kappa\rho^2 H_1 G_2^{(2)} - \alpha\kappa E_2\, G_3^{(2)} + \left[2 - \alpha\rho^2\left(\gamma + \kappa E_1\right)\right]G_4^{(2)}$$
$$- \alpha U_0 G_1^{(1)} + \alpha\rho^2 U G_4^{(1)} = 0, \tag{2.52}$$

$$\frac{\partial G_3^{(2)}}{\rho\partial\rho} - z\frac{\partial G_4^{(2)}}{\partial z} - \alpha\kappa z^2 H_1 G_1^{(2)} + \left[\alpha\left(C_0 + \overline{\varphi}\right) + \alpha\kappa H_2\right]G_2^{(2)} + \alpha\left(\gamma - \kappa E_1\right)G_3^{(2)} - \left(1 - \alpha\kappa z^2 E_2\right)G_4^{(2)}$$
$$- \alpha z^2 U_0 G_2^{(1)} - \alpha z^2 U G_3^{(1)} = 0, \tag{2.53}$$

$$\rho\frac{\partial G_2^{(2)}}{\partial\rho} + z\frac{\partial G_1^{(2)}}{\partial z} + \left(1 + \alpha\kappa z^2 E_2\right)G_1^{(2)} + \left[2 - \alpha\rho^2\left(\gamma - \kappa E_1\right)\right]G_2^{(2)} + \left(2 - \alpha C_0 - \alpha\overline{\varphi} - \alpha\kappa H_2\right)G_3^{(2)}$$
$$- \alpha\kappa\rho^2 z^2 H_1 G_4^{(2)} + \alpha z^2 U_0 G_3^{(1)} + \alpha\rho^2 z^2 U G_2^{(1)} = 0, \tag{2.54}$$

$$\frac{\partial G_1^{(2)}}{\rho\partial\rho} - \frac{\partial G_2^{(2)}}{z\partial z} + \alpha\left(\gamma + \kappa E_1\right)G_1^{(2)} - \alpha\kappa E_2 G_2^{(2)} - \alpha\kappa H_1 G_3^{(2)} + \left(2 - \alpha C_0 - \alpha\overline{\varphi} + \alpha\kappa H_2\right)G_4^{(2)}$$
$$+ \alpha U_0 G_4^{(1)} - \alpha U G_1^{(1)} = 0. \tag{2.55}$$

The components of the vector $\boldsymbol{\Gamma}$ of (2.26) and the ones of the vector $\boldsymbol{\Delta}$ of (2.27) have the structure

$$\Gamma_\rho = \rho\Gamma_1(\rho^2, z^2), \ \Gamma_1 = 2\left(z^2 G_1^{(2)} G_4^{(2)} + G_2^{(2)} G_3^{(2)} - G_1^{(1)} G_4^{(1)} - z^2 G_2^{(1)} G_3^{(1)}\right),$$

$$\Gamma_z = z\Gamma_2(\rho^2, z^2), \ \Gamma_2 = 2\left(\rho^2 G_2^{(1)} G_4^{(1)} - G_1^{(1)} G_3^{(1)} - \rho^2 G_2^{(2)} G_4^{(2)} + G_1^{(2)} G_3^{(2)}\right). \tag{2.56}$$

$$\Delta_\rho = \rho z\Delta_1(\rho^2, z^2), \ \Delta_1 = 2\left(G_3^{(1)} G_4^{(1)} - G_1^{(1)} G_2^{(1)} - G_3^{(2)} G_4^{(2)} + G_1^{(2)} G_2^{(2)}\right), \ \Delta_z = \Delta_2(\rho^2, z^2),$$

$$\Delta_2 = -G_1^{(1)^2} + \rho^2 z^2 G_2^{(1)^2} + z^2 G_3^{(1)^2} - \rho^2 G_4^{(1)^2} + z^2 G_1^{(2)^2} - \rho^2 G_2^{(2)^2} - G_3^{(2)^2} + \rho^2 z^2 G_4^{(2)^2}. \tag{2.57}$$

Now Eqs. (2.38) and (2.39) reduce to

$$\nabla^2\overline{\varphi} + 4\pi\left(G_1^{(1)^2} + \rho^2 z^2 G_2^{(1)^2} + z^2 G_3^{(1)^2} + \rho^2 G_4^{(1)^2} - z^2 G_1^{(2)^2} - \rho^2 G_2^{(2)^2} - G_3^{(2)^2} - \rho^2 z^2 G_4^{(2)^2}\right)$$

$$+ 4\pi\kappa\left(2\Gamma_1 + \Gamma_2 + \rho\frac{\partial\Gamma_1}{\partial\rho} + z\frac{\partial\Gamma_2}{\partial z}\right) = 0\,, \tag{2.58}$$

$$\frac{\partial^2\gamma}{\partial\rho^2} + \frac{3}{\rho}\frac{\partial\gamma}{\partial\rho} + \frac{\partial^2\gamma}{\partial z^2} + 8\pi\left(G_1^{(1)} G_4^{(1)} - z^2 G_2^{(1)} G_3^{(1)} - z^2 G_1^{(2)} G_4^{(2)} + G_2^{(2)} G_3^{(2)}\right)$$

$$- 4\pi\kappa\left(\Delta_1 + z\frac{\partial\Delta_1}{\partial z} - \frac{\partial\Delta_2}{\rho\partial\rho}\right) = 0\,. \tag{2.59}$$

It is worthy of remark that Eqs. (2.48)–(2.55) coincide with Eqs. (4.28)–(4.35) of [2] if one puts $\kappa = 0$ in the former. In the same case Eqs. (2.58)–(2.59) also coincide with Eqs. (4.36)–(4.37) of [2]. The further transformations will be carried out in much the same manner as in Ref. [2]. It follows from (2.40) and (2.41) that

$$G_1^{(2)} = -\rho^2 B_1 G_2^{(2)} - B_2 G_3^{(2)}, \quad G_4^{(2)} = B_2 G_2^{(2)} + B_1 G_3^{(2)}, \tag{2.60}$$

where

$$B_1 = \frac{G_1^{(1)} G_2^{(1)} + G_3^{(1)} G_4^{(1)}}{D}, \quad B_2 = \frac{G_1^{(1)} G_3^{(1)} + \rho^2 G_2^{(1)} G_4^{(1)}}{D}, \quad D = G_1^{(1)^2} - \rho^2 G_4^{(1)^2}. \tag{2.61}$$

These expressions for $G_1^{(2)}$ and $G_4^{(2)}$ can be introduced into all other equations so that the number of the unknown functions reduces.

Solving Eqs. (2.52) and (2.55) simultaneously for $U_0$ and $U$ we obtain

$$\alpha U_0 = \frac{U_1 G_1^{(1)} - \rho^2 U_2 G_4^{(1)}}{D}, \quad \alpha U = \frac{U_1 G_4^{(1)} - U_2 G_1^{(1)}}{D}, \tag{2.62}$$

where

$$U_1 = \rho B_2\frac{\partial G_2^{(2)}}{\partial\rho} + \rho B_1\frac{\partial G_3^{(2)}}{\partial\rho} + \frac{\partial G_3^{(2)}}{z\partial z} + \left[2B_2 + \rho\frac{\partial B_2}{\partial\rho} - \alpha\rho^2(\gamma + \kappa E_1)B_2 - \alpha\rho^2(C_0 + \overline{\varphi} - \kappa H_2)B_1\right.$$

$$\left. - \alpha\kappa\rho^2 H_1\right]G_2^{(2)} + \left[2B_1 + \rho\frac{\partial B_1}{\partial\rho} - \alpha\rho^2(\gamma + \kappa E_1)B_1 - \alpha(C_0 + \overline{\varphi} - \kappa H_2)B_2 - \alpha\kappa E_2\right]G_3^{(2)}, \tag{2.63}$$

$$U_2 = \rho B_1 \frac{\partial G_2^{(2)}}{\partial \rho} + \frac{\partial G_2^{(2)}}{z \partial z} + B_2 \frac{\partial G_3^{(2)}}{\rho \partial \rho} + \left[ 2B_1 + \rho \frac{\partial B_1}{\partial \rho} + \alpha \rho^2 (\gamma + \kappa E_1) B_1 - (2 - \alpha C_0 - \alpha \overline{\varphi} + \alpha \kappa H_2) B_2 \right.$$

$$\left. + \alpha \kappa E_2 \right] G_2^{(2)} + \left[ \frac{\partial B_2}{\rho \partial \rho} + \alpha (\gamma + E_1) B_2 - (2 - \alpha C_0 - \alpha \overline{\varphi} + \alpha \kappa H_2) B_1 + \alpha \kappa H_1 \right] G_3^{(2)}. \qquad (2.64)$$

These $U_0$ and $U$ are to be substituted into all other equations to reduce the number of the unknown functions. In particular, if the substitution is made into Eqs. (2.53) and (2.54), one obtains the following equations

$$\rho C \frac{\partial G_2^{(2)}}{\partial \rho} - 2\rho^2 z B_1 \frac{\partial G_2^{(2)}}{\partial z} + C_{11} G_2^{(2)} + C_{12} G_3^{(2)} = 0, \qquad (2.65)$$

$$C \frac{\partial G_3^{(2)}}{\rho \partial \rho} - 2z B_1 \frac{\partial G_3^{(2)}}{\partial z} + C_{21} G_2^{(2)} + C_{22} G_3^{(2)} = 0, \qquad (2.66)$$

in which

$$C = 1 + z^2 B_3, \quad B_3 = \frac{G_3^{(1)2} - \rho^2 G_2^{(1)2}}{D} = B_2^2 - \rho^2 B_1^2, \qquad (2.67)$$

$$C_{11} = 2 + \rho z^2 \left( B_2 \frac{\partial B_2}{\partial \rho} - \rho^2 B_1 \frac{\partial B_1}{\partial \rho} \right) - \rho^2 z \frac{\partial B_1}{\partial z} - \rho^2 B_1 + 2z^2 B_3 - \alpha \rho^2 \gamma (1 + \rho^2 z^2 B_1^2 + z^2 B_2^2)$$

$$+ 2\rho^2 z^2 (1 - \alpha C_0 - \alpha \overline{\varphi} + \alpha \kappa H_2) B_1 B_2 + \alpha \kappa \rho^2 E_1 (1 - \rho^2 z^2 B_1^2 - z^2 B_2^2) - 2\alpha \kappa \rho^2 z^2 E_2 B_1$$

$$- 2\alpha \kappa \rho^2 z^2 H_1 B_2, \qquad (2.68)$$

$$C_{12} = \rho z^2 \left( B_2 \frac{\partial B_1}{\partial \rho} - B_1 \frac{\partial B_2}{\partial \rho} \right) - z \frac{\partial B_2}{\partial z} - B_2 + 2z^2 B_1 B_2 [1 - \alpha \rho^2 (\gamma + \kappa E_1)] + 2(1 + \rho^2 z^2 B_1^2)$$

$$- \alpha (C_0 + \overline{\varphi})(1 + \rho^2 z^2 B_1^2 + z^2 B_2^2) - 2\alpha \kappa E_2 B_2 - 2\alpha \kappa \rho^2 z^2 H_1 B_1$$

$$- \alpha \kappa H_2 (1 - \rho^2 z^2 B_1^2 - z^2 B_2^2), \qquad (2.69)$$

$$C_{21} = \rho z^2 \left( B_2 \frac{\partial B_1}{\partial \rho} - B_1 \frac{\partial B_2}{\partial \rho} \right) - z \frac{\partial B_2}{\partial z} - B_2 + 2\alpha \rho^2 z^2 (\gamma + \kappa E_1) B_1 B_2 - 2z^2 B_2^2$$

$$+ \alpha (C_0 + \overline{\varphi})(1 + \rho^2 z^2 B_1^2 + z^2 B_2^2) + 2\alpha \kappa z^2 E_2 B_2 + 2\alpha \kappa \rho^2 z^2 H_1 B_1$$

$$+ \alpha \kappa H_2 (1 - \rho^2 z^2 B_1^2 - z^2 B_2^2), \qquad (2.70)$$

$$C_{22} = z^2 \left( B_2 \frac{\partial B_2}{\rho \partial \rho} - \rho B_1 \frac{\partial B_1}{\partial \rho} \right) - z \frac{\partial B_1}{\partial z} - B_1 - 2z^2 B_1^2 + \alpha \gamma (1 + \rho^2 z^2 B_1^2 + z^2 B_2^2)$$

$$- 2z^2 (1 - \alpha C_0 - \alpha \overline{\varphi} + \alpha \kappa H_2) B_1 B_2 - \alpha \kappa E_1 (1 - \rho^2 z^2 B_1^2 - z^2 B_2^2) + 2\alpha \kappa z^2 E_2 B_1$$

$$+ 2\alpha \kappa z^2 H_1 B_2. \qquad (2.71)$$

In regard to the type, Eqs. (2.65)–(2.66) coincide with Eqs. (4.43)–(4.44) of [2] so that they can be treated as in [2].

Equations (2.58) and (2.59) can be solved for $\overline{\varphi}$ and $\gamma$ with use made of (4.50) and (4.51) of [2]. If the resulting $\overline{\varphi}$ and $\gamma$ are placed in the remaining equations, we shall obtain six equations (2.48)–(2.51), (2.65), (2.66) for six functions $G_1^{(1)}, G_2^{(1)}, G_3^{(1)}, G_4^{(1)}, G_2^{(2)}, G_3^{(2)}$ just as in [2]. The equations contain two unknown constants $\kappa$ and $C_0$ which are to be found in the course of solving the equations with the proviso that the resulting functions are nonsingular and vanish sufficiently rapidly at infinity.

As in Ref. [2], one can deduce second-order differential equations instead of Eqs. (2.48)–(2.51):

$$\nabla^2 F_1^{(1)} - \nu^2 F_1^{(1)} + Q_1 = 0, \ \nabla^2 F_2^{(1)} - \frac{F_2^{(1)}}{\rho^2} - \nu^2 F_2^{(1)} + \rho z Q_2 = 0,$$

$$\nabla^2 F_3^{(1)} - \nu^2 F_3^{(1)} + z Q_3 = 0, \ \nabla^2 F_4^{(1)} - \frac{F_4^{(1)}}{\rho^2} - \nu^2 F_4^{(1)} + \rho Q_4 = 0, \tag{2.72}$$

where

$$\nu^2 = \alpha C_0 (2 - \alpha C_0), \tag{2.73}$$

$$Q_1 = -\rho \frac{\partial K_4^{(1)}}{\partial \rho} - z \frac{\partial K_3^{(1)}}{\partial z} - 2K_4^{(1)} - K_3^{(1)} - (2 - \alpha C_0) K_1^{(1)}, \ Q_2 = -\frac{\partial K_3^{(1)}}{\rho \partial \rho} + \frac{\partial K_4^{(1)}}{z \partial z} - (2 - \alpha C_0) K_2^{(1)},$$

$$Q_3 = \rho \frac{\partial K_2^{(1)}}{\partial \rho} + \frac{\partial K_1^{(1)}}{z \partial z} + 2K_2^{(1)} + \alpha C_0 K_3^{(1)}, \ Q_4 = \frac{\partial K_1^{(1)}}{\rho \partial \rho} - z \frac{\partial K_2^{(1)}}{\partial z} - K_2^{(1)} + \alpha C_0 K_4^{(1)}, \tag{2.74}$$

and

$$K_1^{(1)} = \alpha (\overline{\varphi} - \kappa H_2) G_1^{(1)} - \alpha \kappa \rho^2 z^2 H_1 G_2^{(1)} - \alpha \kappa z^2 E_2 G_3^{(1)} - \alpha \rho^2 (\gamma + E_1) G_4^{(1)}$$
$$+ \alpha z^2 U_0 G_1^{(2)} - \alpha \rho^2 z^2 U G_4^{(2)}, \tag{2.75}$$

$$K_2^{(1)} = -\alpha \kappa H_1 G_1^{(1)} + \alpha (\overline{\varphi} + \kappa H_2) G_2^{(1)} + \alpha (\gamma - \kappa E_1) G_3^{(1)} + \alpha \kappa E_2 G_4^{(1)}$$
$$+ \alpha U_0 G_2^{(2)} + \alpha U G_3^{(2)}, \tag{2.76}$$

$$K_3^{(1)} = -\alpha \kappa E_2 G_1^{(1)} + \alpha \rho^2 (\gamma - \kappa E_1) G_2^{(1)} + \alpha (\overline{\varphi} + \kappa H_2) G_3^{(1)} + \alpha \kappa \rho^2 H_1 G_4^{(1)}$$
$$+ \alpha U_0 G_3^{(2)} + \alpha \rho^2 U G_2^{(2)}, \tag{2.77}$$

$$K_4^{(1)} = -\alpha (\gamma + \kappa E_1) G_1^{(1)} + \alpha \kappa z^2 E_2 G_2^{(1)} + \alpha \kappa z^2 H_1 G_3^{(1)} + \alpha (\overline{\varphi} - \kappa H_2) G_4^{(1)}$$
$$+ \alpha z^2 U_0 G_4^{(2)} - \alpha z^2 U G_1^{(2)}. \tag{2.78}$$

The equations of (2.72) can be rewritten in an integral form given by Eqs. (4.59) and (A.5) of [2] from which it follows that the behavior of the functions in question as $r \to \infty$ is determined first of all by the factor $e^{-\nu r}$ (see also [2]). Therefore the density $\rho_m$ of (2.18) is proportional to $e^{-2\nu r}$, which shows that the proton radius is approximately equal to $(1/2\nu)\lambdabar_p$ for the unit of length in the present case is the proton Compton wavelength $\lambdabar_p \approx 0.21$ fm. Experiment indicates that the proton radius is about $4\,\lambdabar_p$ [6]. Consequently, $\nu \approx 1/8$ for the proton. We are now in position to evaluate the constant $C_0$ that figures in the above equations. Taking the smaller root in Eq. (2.73), since $C_0$ should be small compared to $1/\alpha \approx 137$, we have $C_0 \approx 1.07$ ($\alpha C_0 \approx 0.0078$).

The value of $\lambdabar_p$ used above is approximate. The exact relation between $\lambdabar_p$ and the proton mass $m_p$ measured experimentally can be found by analogy with Eq. (5.2) of [2]. One should substitute $\mathscr{E} = m_p c^2$ into Eq. (2.47), which yields

$$\lambdabar_p = \frac{\hbar}{m_p c}\left\{1 - \alpha C_0 + \alpha \int\limits_{(\infty)}\left[\frac{1}{8\pi}\left(\widetilde{E}^2 + \widetilde{H}^2\right) - \overline{\varphi}\left(\widetilde{\Psi}_1^*\widetilde{\Psi}_1 - \widetilde{\Psi}_2^*\widetilde{\Psi}_2\right) - \kappa\widetilde{\mathbf{E}}\widetilde{\boldsymbol{\Gamma}}\right]d\widetilde{V}\right\}. \tag{2.79}$$

If we put $\alpha = 0$ here, we obtain $\lambdabar_p = \hbar/m_p c$, the relation usually used to calculate $\lambdabar_p$.

The equations obtained above can have not only the solution relevant to the proton (antiproton) but also solutions of higher energy $\mathscr{E} = \int T^{00}dV$ that must correspond to other baryons. Upon knowing the energy spectrum one will be able to calculate the baryon mass spectrum by $m = \mathscr{E}/c^2$. The heavier baryons should be metastable because they correspond to excited states of the protonic field.

## 2.2. Non-relativistic limit

The passage to the non-relativistic limit in the present theory is explained in Sec. 5 of [2]. We should neglect the self-field by setting $A_\mu = V_\mu = 0$ in Eq. (2.4) but the external electromagnetic field (if it exists) should be retained instead. Besides, we should introduce the proton mass $m_p$ according to $\lambdabar_p = \hbar/m_p c$ (by the way, this follows from (2.79) when $\alpha = e^2/\hbar c \to 0$). Similarly to (2.21) we write down the resulting equation in the three-dimensional notation:

$$i\hbar\frac{\partial\Psi_1}{\partial t} + ic\hbar\boldsymbol{\alpha}\nabla\Psi_1 - e\mathbf{A}^{\text{ext}}\boldsymbol{\alpha}\Psi_1 + e\varphi^{\text{ext}}\Psi_1 - m_p c^2\beta\Psi_1 + ie\chi\left(\mathbf{E}^{\text{ext}}\boldsymbol{\gamma} + \mathbf{H}^{\text{ext}}\boldsymbol{\beta}\right)\Psi_1 = 0. \tag{2.80}$$

Henceforth, we shall omit the superscript "ext" for brevity sake.

The subsequent passage to the limit $c \to \infty$ is carried out in much the same way as in § 33 of [4]. First, we exclude the proton rest energy $m_p c$ with the help of the substitution $\Psi_1 = \psi\exp(-im_p c^2 t/\hbar)$. Secondly, we represent the bispinor $\psi$ as $\psi = \begin{pmatrix} \psi' \\ \psi'' \end{pmatrix}$ and obtain two equations for $\psi'$ and $\psi''$. In the second equation, only the quantity $2m_p c^2$ in the coefficient of $\psi''$ is retained so that

$$\psi'' = \frac{\boldsymbol{\sigma}}{2m_p c}\left(-i\hbar\nabla + \frac{e}{c}\mathbf{A} + \frac{ie}{c}\chi\mathbf{E}\right)\psi' .\tag{2.81}$$

Substituting this into the first equation and performing transformations as in § 33 of [4] we arrive at

$$i\hbar\frac{\partial\psi'}{\partial t} = \left\{\frac{1}{2m_p}\left(-i\hbar\nabla + \frac{e}{c}\mathbf{A}\right)^2 + \frac{e\hbar}{2m_p c}\boldsymbol{\sigma}\mathbf{H} + e\chi\boldsymbol{\sigma}\mathbf{H} - e\varphi - \frac{e\chi}{2m_p c}\left(\hbar\,\mathrm{div}\mathbf{E} - \frac{e\chi}{c}E^2\right)\right.$$
$$\left. - \frac{e\chi}{2m_p c}\boldsymbol{\sigma}\left(i\hbar\,\mathrm{curl}\mathbf{E} + 2\left[\mathbf{E}\left(-i\hbar\nabla + \frac{e}{c}\mathbf{A}\right)\right]\right)\right\}\psi' .\tag{2.82}$$

The coefficient of $\mathbf{H}$ indicates that the proton has the magnetic moment

$$\boldsymbol{\mu} = -\frac{e\hbar}{2m_p c}\left(1 + \frac{2m_p c}{\hbar}\chi\right)\boldsymbol{\sigma} .\tag{2.83}$$

It will be recalled that the charge of the proton is $-e > 0$ so that the magnetic moment is oriented in the direction of the spin as it should be. If we introduce the nuclear magneton $\mu_N = |e|\hbar/(2m_p c)$ and replace $\chi$ by $\kappa$ according to (2.20), the value of the magnetic moment will be $\mu = (1 + 2\kappa)\mu_N$. The experimental value of the proton magnetic moment is $\mu \approx 2.793\ \mu_N$. Therefore, $\kappa \approx 0.895$. The last value is approximate for Eq. (2.83) is approximate.

### 3. Simultaneous description of the electronic and protonic fields

When considering simultaneously the electronic and protonic fields the question arises as to whether or not the four-vector $V_\mu$ (or $v_\mu$ in [2]) is common to the electronic and protonic fields. In Ref. [2] where the vector $v_\mu$ was introduced it came from the imaginary part of the four-potential $A_\mu$. Seeing that the potential $A_\mu$ is unique in space it would seem that $v_\mu$ should be unique for both the fields as well. This is not the case, however. The uniqueness of $v_\mu$ would signify that there is a new and unknown field in nature represented by the vector $v_\mu$, the field that is not described by any equations. Besides, the vector $V_\mu$ does not figure in the energy-momentum tensor of (2.14). All of this amounts to saying that the electronic field is characterized by $\psi_1$, $\psi_2$ and $v_\mu$ while the protonic by $\Psi_1$, $\Psi_2$ and $V_\mu$. We shall yet return to this question a little later.

When the electronic field was considered in Ref. [2], as the standard of length was chosen the electron Compton wavelength $\lambdabar$. In Sec. 2 of the present paper the standard of length for the protonic field was the proton Compton wavelength $ƛ_p$. When treating simultaneously the fields the standard of length must be unique. In the present section our main concern will be nuclei. For this reason the standard of length will be $ƛ_p$. As for the electron Compton wavelength $ƛ$ we shall write $ƛ = \zeta ƛ_p$ where experimentally $\zeta \approx 1840$. The exact value of $\zeta$ is to be found in the course of solving equations obtained below. The knowledge of $\zeta$ enables one to find out the relation between the electron mass $m$ and the proton mass $m_p$ with use made of Eq. (5.2) of [2] and of Eq. (2.79) above. In this way, one of the fundamental constants in nature, the ratio $m_p/m$, will be explained.

Now we combine the Lagrangian of (3.3) of [2] where the replacement $m = \hbar/ƛc = \hbar/(\zeta ƛ_p c)$ is made and the Lagrangian of (2.3):

$$L(x) = -\frac{1}{16\pi}F_{\mu\nu}F^{\mu\nu} + \frac{ic\hbar}{2}\left(\overline{\psi}_1\gamma^\mu\frac{\partial\psi_1}{\partial x^\mu} - \frac{\partial\overline{\psi}_1}{\partial x^\mu}\gamma^\mu\psi_1 - \overline{\psi}_2\gamma^\mu\frac{\partial\psi_2}{\partial x^\mu} + \frac{\partial\overline{\psi}_2}{\partial x^\mu}\gamma^\mu\psi_2\right)$$

$$+\frac{ic\hbar}{2}\left(\overline{\Psi}_1\gamma^\mu\frac{\partial\Psi_1}{\partial x^\mu} - \frac{\partial\overline{\Psi}_1}{\partial x^\mu}\gamma^\mu\Psi_1 - \overline{\Psi}_2\gamma^\mu\frac{\partial\Psi_2}{\partial x^\mu} + \frac{\partial\overline{\Psi}_2}{\partial x^\mu}\gamma^\mu\Psi_2\right) - eA_\mu(\overline{\psi}_1\gamma^\mu\psi_1 - \overline{\psi}_2\gamma^\mu\psi_2)$$

$$+ eA_\mu(\overline{\Psi}_1\gamma^\mu\Psi_1 - \overline{\Psi}_2\gamma^\mu\Psi_2) + \frac{i}{2}e\chi F_{\mu\nu}(\overline{\Psi}_1\sigma^{\mu\nu}\Psi_1 - \overline{\Psi}_2\sigma^{\mu\nu}\Psi_2 - \frac{c\hbar}{\zeta ƛ_p}(\overline{\psi}_1\psi_1 - \overline{\psi}_2\psi_2)$$

$$-\frac{c\hbar}{ƛ_p}(\overline{\Psi}_1\Psi_1 - \overline{\Psi}_2\Psi_2) - ev_\mu(\overline{\psi}_1\gamma^\mu\psi_2 + \overline{\psi}_2\gamma^\mu\psi_1) + eV_\mu(\overline{\Psi}_1\gamma^\mu\Psi_2 + \overline{\Psi}_2\gamma^\mu\Psi_1)\,. \quad (3.1)$$

The Lagrangian is gauge invariant according to (2.2).

Some comment at this point is in order. The Lagrangian of (3.1) can be supplemented with terms of the type $\psi_1\Psi_1$ that would reflect direct interaction of the electronic and protonic fields. In this case, however, the equations of [2] describing the electronic field will be supplemented with terms containing $\Psi_{1,2}$ and relevant to the protonic field, which contradicts equations of QED, and the equations of Sec. 2 above will be supplemented with terms containing $\psi_{1,2}$ and relevant to the electronic field. And what is more important, those terms will make it possible for the proton to be converted into a positron and vice versa, which is not observed experimentally up till now. We do not include such terms in the Lagrangian of (3.1) presuming that the electronic and protonic fields interact with each other only via the electromagnetic field.

Varying $L(x)$ with respect to $\overline{\psi}_1$, $\overline{\psi}_2$, $\psi_1$ and $\psi_2$ we obtain Eqs. (3.4)–(3.7) of [2] with the above replacement $m = \hbar/(\zeta ƛ_p c)$ while variation with respect to $\overline{\Psi}_1$, $\overline{\Psi}_2$, $\Psi_1$ and $\Psi_2$ leads to Eqs. (2.4)–(2.7) above. Only the equation stemming from variation with respect to $A_\mu$ changes:

$$\frac{\partial F^{\mu\nu}}{\partial x^{\nu}} = -4\pi e\left(\overline{\psi}_1\gamma^{\mu}\psi_1 - \overline{\psi}_2\gamma^{\mu}\psi_2\right) + 4\pi e\left(\overline{\Psi}_1\gamma^{\mu}\Psi_1 - \overline{\Psi}_2\gamma^{\mu}\Psi_2\right) + 4\pi ie\chi\,\frac{\partial}{\partial x^{\nu}}\left(\overline{\Psi}_1\sigma^{\mu\nu}\Psi_1 - \overline{\Psi}_2\sigma^{\mu\nu}\Psi_2\right).$$

(3.2)

Varying $L(x)$ with respect to $v_{\mu}$ and $V_{\mu}$ one obtains

$$\overline{\psi}_1\gamma^{\mu}\psi_2 + \overline{\psi}_2\gamma^{\mu}\psi_1 = 0, \ \ \overline{\Psi}_1\gamma^{\mu}\Psi_2 + \overline{\Psi}_2\gamma^{\mu}\Psi_1 = 0\,.$$

(3.3)

In this connection the following remark may be made. If the four-vector $v_{\mu}$ were unique for both the electronic and protonic fields as discussed at the outset of this section, instead of the two equations of (3.3) one would have one equation:

$$\overline{\psi}_1\gamma^{\mu}\psi_2 + \overline{\psi}_2\gamma^{\mu}\psi_1 + \overline{\Psi}_1\gamma^{\mu}\Psi_2 + \overline{\Psi}_2\gamma^{\mu}\Psi_1 = 0\,.$$

(3.4)

In the case of a hydrogen atom the protonic field is confined in a region of size several $\lambdabar_{\mathrm{p}}$ while the electronic field extends to the Bohr radius $r_{\mathrm{B}} = m_{\mathrm{p}}\lambdabar_{\mathrm{p}}/(\alpha m) \approx 2.5{\cdot}10^5\ \lambdabar_{\mathrm{p}}$. Therefore, the region where the first two terms in (3.4) are essential and the region where the last two terms are essential are substantially different. Consequently, only the first two terms in (3.4) can compensate each other, which leads to the first equation of (3.3), and the mutual compensation of the last two terms in (3.4) yields the second equation of (3.3).

From Eqs. (3.4)–(3.7) of [2] and from Eqs. (2.4)–(2.7) above it follows that

$$\frac{\partial}{\partial x^{\mu}}\left(\overline{\psi}_1\gamma^{\mu}\psi_1 - \overline{\psi}_2\gamma^{\mu}\psi_2\right) = 0, \ \ \frac{\partial}{\partial x^{\mu}}\left(\overline{\Psi}_1\gamma^{\mu}\Psi_1 - \overline{\Psi}_2\gamma^{\mu}\Psi_2\right) = 0\,.$$

(3.5)

It should be observed that, if we differentiate (3.2) with respect to $x^{\mu}$ and take the antisymmetry of $F^{\mu\nu}$ and $\sigma^{\mu\nu}$ into account, the resulting equation will hold owing to (3.5). In perfect analogy to (2.11) we shall obtain from the two equations of (3.5) that

$$\int\limits_{(\infty)}(\psi_1^*\psi_1 - \psi_2^*\psi_2)dV = N_{\mathrm{e}}, \ \ \int\limits_{(\infty)}(\Psi_1^*\Psi_1 - \Psi_2^*\Psi_2)dV = N_{\mathrm{p}}\,,$$

(3.6)

where $N_{\mathrm{e}}$ and $N_{\mathrm{p}}$ are the number of electrons and protons respectively in the system. Inasmuch as the integrals in (3.6) are constants of motion, these numbers are conserved in all processes described by the above equations.

We present also the energy-momentum tensor which can be calculated starting from (3.1) similarly to (2.14):

$$T^{\mu\nu} = \frac{1}{16\pi}\left(g^{\mu\nu}F_{\lambda\eta}F^{\lambda\eta} - 4F^{\mu\lambda}F^{\nu}{}_{\lambda}\right) - \frac{e}{2}\left(A^{\mu}\overline{\psi}_1\gamma^{\nu}\psi_1 + A^{\nu}\overline{\psi}_1\gamma^{\mu}\psi_1 - A^{\mu}\overline{\psi}_2\gamma^{\nu}\psi_2 - A^{\nu}\overline{\psi}_2\gamma^{\mu}\psi_2\right)$$

$$+ \frac{e}{2}\left(A^{\mu}\overline{\Psi}_1\gamma^{\nu}\Psi_1 + A^{\nu}\overline{\Psi}_1\gamma^{\mu}\Psi_1 - A^{\mu}\overline{\Psi}_2\gamma^{\nu}\Psi_2 - A^{\nu}\overline{\Psi}_2\gamma^{\mu}\Psi_2\right)$$

$$- \frac{ie\chi}{2}\left(F^{\mu}{}_{\eta}\overline{\Psi}_1\sigma^{\eta\nu}\Psi_1 + F^{\nu}{}_{\eta}\overline{\Psi}_1\sigma^{\eta\mu}\Psi_1 - F^{\mu}{}_{\eta}\overline{\Psi}_2\sigma^{\eta\nu}\Psi_2 - F^{\nu}{}_{\eta}\overline{\Psi}_2\sigma^{\eta\mu}\Psi_2\right)$$

$$+ \frac{ich}{4} \left( \overline{\psi}_1 \gamma^\nu \frac{\partial \psi_1}{\partial x_\mu} + \overline{\psi}_1 \gamma^\mu \frac{\partial \psi_1}{\partial x_\nu} - \frac{\partial \overline{\psi}_1}{\partial x_\mu} \gamma^\nu \psi_1 - \frac{\partial \overline{\psi}_1}{\partial x_\nu} \gamma^\mu \psi_1 - \overline{\psi}_2 \gamma^\nu \frac{\partial \psi_2}{\partial x_\mu} - \overline{\psi}_2 \gamma^\mu \frac{\partial \psi_2}{\partial x_\nu} \right.$$

$$\left. + \frac{\partial \overline{\psi}_2}{\partial x_\mu} \gamma^\nu \psi_2 + \frac{\partial \overline{\psi}_2}{\partial x_\nu} \gamma^\mu \psi_2 \right) + \frac{ich}{4} \left( \overline{\Psi}_1 \gamma^\nu \frac{\partial \Psi_1}{\partial x_\mu} + \overline{\Psi}_1 \gamma^\mu \frac{\partial \Psi_1}{\partial x_\nu} - \frac{\partial \overline{\Psi}_1}{\partial x_\mu} \gamma^\nu \Psi_1 - \frac{\partial \overline{\Psi}_1}{\partial x_\nu} \gamma^\mu \Psi_1 \right.$$

$$\left. - \overline{\Psi}_2 \gamma^\nu \frac{\partial \Psi_2}{\partial x_\mu} - \overline{\Psi}_2 \gamma^\mu \frac{\partial \Psi_2}{\partial x_\nu} + \frac{\partial \overline{\Psi}_2}{\partial x_\mu} \gamma^\nu \Psi_2 + \frac{\partial \overline{\Psi}_2}{\partial x_\nu} \gamma^\mu \Psi_2 \right), \tag{3.7}$$

and the vector **M** of angular momentum as in (2.17)

$$\mathbf{M} = \frac{1}{4\pi c} \int\limits_{(\infty)} [\mathbf{r}[\mathbf{EH}]] dV - \frac{e}{c} \int\limits_{(\infty)} \left( [\mathbf{rA}](\overline{\psi}_1 \gamma^0 \psi_1 - \overline{\psi}_2 \gamma^0 \psi_2 - \overline{\Psi}_1 \gamma^0 \Psi_1 + \overline{\Psi}_2 \gamma^0 \Psi_2) + \chi[\mathbf{r}[\mathbf{H\Gamma}]] \right) dV$$

$$+ \hbar \int\limits_{(\infty)} \left\{ \frac{1}{2} \left( \overline{\psi}_1 \gamma^0 \mathbf{\Sigma} \psi_1 - \overline{\psi}_2 \gamma^0 \mathbf{\Sigma} \psi_2 \right) - i \left( \overline{\psi}_1 \gamma^0 [\mathbf{r}\nabla \psi_1] - \overline{\psi}_2 \gamma^0 [\mathbf{r}\nabla \psi_2] \right) \right\} dV$$

$$+ \hbar \int\limits_{(\infty)} \left\{ \frac{1}{2} \left( \overline{\Psi}_1 \gamma^0 \mathbf{\Sigma} \Psi_1 - \overline{\Psi}_2 \gamma^0 \mathbf{\Sigma} \Psi_2 \right) - i \left( \overline{\Psi}_1 \gamma^0 [\mathbf{r}\nabla \Psi_1] - \overline{\Psi}_2 \gamma^0 [\mathbf{r}\nabla \Psi_2] \right) \right\} dV. \tag{3.8}$$

In Ref. [3] it was stated that several electrons are described by a single electronic field where there are no individual electrons. This suggests that the protonic field should also describe not only one proton but several protons as well. The numbers of electrons and protons in a system are given by (3.6). It was argued in Introduction that all elementary particles are to be describable in terms of the electronic and protonic fields. This holds for nuclei too. Figuratively speaking, they are composed of electrons and protons. As mentioned in Introduction this was, historically, the first hypothesis concerning the structure of the nuclei. The hypothesis was rejected because of a naïve consideration of the spins. We shall demonstrate below that the spin of the nuclei can be explained in terms of the electronic and protonic fields.

If the electronic field exists in a nucleus, it exists of course in the electronic envelope of the atom formed by the same nucleus. The question arises as to whether the electronic field of the nucleus and the one of the electronic envelope are one and the same field. The electronic field of the nucleus cannot vanish identically in the electronic envelope, and conversely the field of the electronic envelope penetrates into the nucleus. Therefore, both the fields are in fact a single electronic field. The atom can be completely ionized, in which case there will remain the nuclear electronic field alone. On the other hand, there are no nuclei composed completely of protons excluding hydrogen [1]H. Therefore all nuclei (except for [1]H) contain the electronic field. As long as the electronic field is unique, any change in the electronic envelope of the atom will affect the atomic nucleus. Until the present time it was believed that the atomic nucleus and the electronic envelope of the atom are independent of each other so that changes in the electronic envelope have no influence on the atomic nucleus.

### 3.1. Dimensionless equations

We introduce dimensionless quantities, denoting them by a tilde, according to (2.20). The dimensionless quantities for the electronic field will be

$$\psi_{1,2} = \frac{1}{\tilde{\lambda}_p^{3/2}} \widetilde{\psi}_{1,2}, \quad v_\mu = \frac{e}{\tilde{\lambda}_p} \widetilde{v}_\mu. \tag{3.9}$$

We shall use the three-dimensional notation. Equations (3.4)–(3.5) of [2] that follow from the Lagrangian of (3.1) become

$$i \frac{\partial \widetilde{\psi}_1}{\partial \widetilde{t}} + i\boldsymbol{\alpha} \widetilde{\nabla} \widetilde{\psi}_1 - \alpha \widetilde{\mathbf{A}} \boldsymbol{\alpha} \widetilde{\psi}_1 + \alpha \widetilde{\varphi} \widetilde{\psi}_1 - \frac{1}{\varsigma} \beta \widetilde{\psi}_1 - \alpha \widetilde{v}_0 \widetilde{\psi}_2 + \alpha \widetilde{\mathbf{v}} \boldsymbol{\alpha} \widetilde{\psi}_2 = 0 \,, \tag{3.10}$$

$$i \frac{\partial \widetilde{\psi}_2}{\partial \widetilde{t}} + i\boldsymbol{\alpha} \widetilde{\nabla} \widetilde{\psi}_2 - \alpha \widetilde{\mathbf{A}} \boldsymbol{\alpha} \widetilde{\psi}_2 + \alpha \widetilde{\varphi} \widetilde{\psi}_2 - \frac{1}{\varsigma} \beta \widetilde{\psi}_2 + \alpha \widetilde{v}_0 \widetilde{\psi}_1 - \alpha \widetilde{\mathbf{v}} \boldsymbol{\alpha} \widetilde{\psi}_1 = 0 \,. \tag{3.11}$$

It should be noted that these equations differ from Eqs. (4.2)–(4.3) of [2] in the sign in front of **A** and φ because the proton charge is taken now as a basis. Besides, the fifth terms in (3.10)–(3.11) contain the factor $1/\varsigma$. Equations (2.21)–(2.22) remain as they stand. In the stationary state considered below Eq. (3.2) yields two equations

$$\nabla^2 \widetilde{\varphi} - 4\pi \big( \widetilde{\psi}_1^* \widetilde{\psi}_1 - \widetilde{\psi}_2^* \widetilde{\psi}_2 \big) + 4\pi \big( \widetilde{\Psi}_1^* \widetilde{\Psi}_1 - \widetilde{\Psi}_2^* \widetilde{\Psi}_2 \big) + 4\pi\kappa \mathrm{div} \widetilde{\boldsymbol{\Gamma}} = 0 \,, \tag{3.12}$$

$$\nabla^2 \widetilde{\mathbf{A}} - \nabla \mathrm{div} \widetilde{\mathbf{A}} - 4\pi \big( \widetilde{\psi}_1^* \boldsymbol{\alpha} \widetilde{\psi}_1 - \widetilde{\psi}_2^* \boldsymbol{\alpha} \widetilde{\psi}_2 \big) + 4\pi \big( \widetilde{\Psi}_1^* \boldsymbol{\alpha} \widetilde{\Psi}_1 - \widetilde{\Psi}_2^* \boldsymbol{\alpha} \widetilde{\Psi}_2 \big) - 4\pi\kappa \mathrm{curl} \widetilde{\mathbf{\Delta}} = 0 \,. \tag{3.13}$$

We look for stationary solutions. In Ref. [2] and in Sec. 2 above we neglected the time-dependent factor $\exp(-i\varepsilon t/\hbar)$ that figured in the wave function in this instance upon assuming that the energy ε was added to the unknown constant part of φ. The present situation is more involved as long as the time-dependent factor is different in $\psi_{1,2}$ and $\Psi_{1,2}$. For this reason we proceed as follows. First of all, we separate out the constant part in the potential φ upon writing

$$\widetilde{\varphi} = \overline{C}_0 + \overline{\varphi}(\mathbf{r}) \,, \tag{3.14}$$

wherein $\overline{\varphi}(\mathbf{r}) \to 0$ as $r \to \infty$. The time-dependent factor in $\psi_{1,2}$ is $\exp(-i\varepsilon_e t/\hbar)$. The electron energy can be written as $\varepsilon_e = mc^2 \widetilde{\varepsilon}_e = (\hbar c/\tilde{\lambda}) \widetilde{\varepsilon}_e = (\hbar c/\varsigma \tilde{\lambda}_p) \widetilde{\varepsilon}_e$. Substituting this and (3.14) into the diagonal terms in (3.10) (the first, fourth and fifth terms) and recalling that the matrix β in the standard representation is $\beta = \begin{pmatrix} 1 & 0 \\ 0 & -1 \end{pmatrix}$ [4] we obtain, in the first two equations of (3.10), $\big[ \alpha C_e + \alpha \overline{\varphi} \big] \widetilde{\psi}_1$ where $\alpha C_e = (\widetilde{\varepsilon}_e - 1)/\varsigma + \alpha \overline{C}_0$. In the last two equations of (3.10) we have $\big[ (\widetilde{\varepsilon}_e + 1)/\varsigma + \alpha \overline{\varphi} \big] \widetilde{\psi}_1 = \big[ 2/\varsigma + \alpha C_e + \alpha \overline{\varphi} \big] \widetilde{\psi}_1$. The time-dependent factor in $\Psi_{1,2}$ is $\exp(-i\varepsilon_p t/\hbar)$ where $\varepsilon_p = m_p c^2 \widetilde{\varepsilon}_p = (\hbar c/\tilde{\lambda}_p) \widetilde{\varepsilon}_p$. Substituting this and (3.14) into the diagonal terms in (2.21) we obtain, in the first two equations of (2.21), $[-\alpha C_p - \alpha\overline{\varphi}]\tilde{\Psi}_1$ where $\alpha C_p = -\tilde{\varepsilon}_p + 1 + \alpha\overline{C}_0$. In the last two equations of (2.21) we have $[\tilde{\varepsilon}_p + 1 - \alpha\overline{\varphi}]\tilde{\Psi}_1 = [2 - \alpha C_p - \alpha\overline{\varphi}]\tilde{\Psi}_1$. Comparing the above formulae with equations of [2] and of Sec. 2 above we can see that the constants $C_e$ and $C_p$ should be small versus $1/\alpha$. It may be added that the constant $\overline{C}_0$ that figures in (3.14) disappears off the equations, which is natural because the potential is defined up to an arbitrary constant. From here on we shall omit the tilde as in Sec. 2.

In what follows we shall restrict ourselves to axially symmetric solutions. Analogously to (2.25) the solution to the above equations in the cylindrical coordinates $\rho$, $\dot{\varphi}$, $z$ will be sought in the form

$$\psi_{1,2} = \begin{pmatrix} f_1^{(1,2)}(\rho, z)e^{ia\dot{\varphi}} \\ f_2^{(1,2)}(\rho, z)e^{i\dot{\varphi}+ia\dot{\varphi}} \\ if_3^{(1,2)}(\rho, z)e^{ia\dot{\varphi}} \\ if_4^{(1,2)}(\rho, z)e^{i\dot{\varphi}+ia\dot{\varphi}} \end{pmatrix}, \qquad \Psi_{1,2} = \begin{pmatrix} F_1^{(1,2)}(\rho, z)e^{ib\dot{\varphi}} \\ F_2^{(1,2)}(\rho, z)e^{i\dot{\varphi}+ib\dot{\varphi}} \\ iF_3^{(1,2)}(\rho, z)e^{ib\dot{\varphi}} \\ iF_4^{(1,2)}(\rho, z)e^{i\dot{\varphi}+ib\dot{\varphi}} \end{pmatrix}, \qquad (3.15)$$

in which the functions $f_i^{(1,2)}$ and $F_i^{(1,2)}$ are real. Compared to Eq. (4.4) of [2] and Eq. (2.25) above, here we have added factors containing the constants $a$ and $b$. It would seem that such a factor could be eliminated by the gauge transformation of (2.2). The factors, however, are different if $a \neq b$ and cannot be eliminated simultaneously unless $a = -b$. Besides, if, for example, $f \propto a\dot{\varphi}$ in (2.2), the $\dot{\varphi}$-component of the vector $\mathbf{A}$ will contain a singular term proportional to $a/\rho$ whereas we always presume that the vector potential $\mathbf{A}$ is regular in all space.

Provided the bispinors $\psi_{1,2}$ and $\Psi_{1,2}$ are given by (3.15), the current density will have only one nonzero component as in (2.28). The vector potential $\mathbf{A}$ will have only one nonzero component $A_{\dot{\varphi}}$ as well (denoted henceforth as $A$). The vectors $\mathbf{v}$ and $\mathbf{V}$ will have only one nonzero components $v_{\dot{\varphi}} = v$ and $V_{\dot{\varphi}} = V$, too.

By way of example we consider Eq. (3.10) into which (3.15) is substituted. When the diagonal terms are transformed as pointed out above, we obtain

$$\frac{\partial f_4^{(1)}}{\partial \rho} + \frac{\partial f_3^{(1)}}{\partial z} + \frac{a+1}{\rho}f_4^{(1)} + \alpha A f_4^{(1)} - \alpha(C_e + \overline{\varphi})f_1^{(1)} + \alpha v_0 f_1^{(2)} - \alpha v f_4^{(2)} = 0, \qquad (3.16)$$

$$\frac{\partial f_3^{(1)}}{\partial \rho} - \frac{\partial f_4^{(1)}}{\partial z} - \frac{a}{\rho}f_3^{(1)} - \alpha A f_3^{(1)} - \alpha(C_e + \overline{\varphi})f_2^{(1)} + \alpha v_0 f_2^{(2)} + \alpha v f_3^{(2)} = 0, \qquad (3.17)$$

$$\frac{\partial f_2^{(1)}}{\partial \rho} + \frac{\partial f_1^{(1)}}{\partial z} + \frac{a+1}{\rho}f_2^{(1)} + \alpha A f_2^{(1)} + (2/\varsigma + \alpha C_e + \alpha\overline{\varphi})f_3^{(1)} - \alpha v_0 f_3^{(2)} - \alpha v f_2^{(2)} = 0, \qquad (3.18)$$

$$\frac{\partial f_1^{(1)}}{\partial \rho} - \frac{\partial f_2^{(1)}}{\partial z} - \frac{a}{\rho} f_1^{(1)} - \alpha A f_1^{(1)} + (2/\varsigma + \alpha C_e + \alpha \overline{\varphi}) f_4^{(1)} - \alpha v_0 f_4^{(2)} + \alpha v f_1^{(2)} = 0. \tag{3.19}$$

In all these equations there are singular terms containing $1/\rho$. In order to get rid of them we make the replacement

$$f_i^{(1,2)} = \rho^\eta \widehat{f}_i^{(1,2)}, \tag{3.20}$$

presuming that the functions $\widehat{f}_i^{(1,2)}$ are regular. An important remark should be made at this point. All physical quantities are quadratic in $f_i^{(1,2)}$ and thereby they will be proportional to $\rho^{2\eta}$ according to (3.20). The physical quantities, their derivatives inclusive, cannot be singular anywhere, including the axis $\rho = 0$. Therefore the number $2\eta$ must be a positive integer or zero. Consequently, the number $\eta$ must be a positive integer, zero or a positive half-integer.

Inserting (3.20) into Eqs. (3.16)–(3.19) gives

$$\frac{\partial \widehat{f}_4^{(1)}}{\partial \rho} + \frac{\partial \widehat{f}_3^{(1)}}{\partial z} + \frac{\eta + a + 1}{\rho} \widehat{f}_4^{(1)} + \alpha A \widehat{f}_4^{(1)} - \alpha(C_e + \overline{\varphi}) \widehat{f}_1^{(1)} + \alpha v_0 \widehat{f}_1^{(2)} - \alpha v \widehat{f}_4^{(2)} = 0, \tag{3.21}$$

$$\frac{\partial \widehat{f}_3^{(1)}}{\partial \rho} - \frac{\partial \widehat{f}_4^{(1)}}{\partial z} + \frac{\eta - a}{\rho} \widehat{f}_3^{(1)} - \alpha A \widehat{f}_3^{(1)} - \alpha(C_e + \overline{\varphi}) \widehat{f}_2^{(1)} + \alpha v_0 \widehat{f}_2^{(2)} + \alpha v \widehat{f}_3^{(2)} = 0, \tag{3.22}$$

$$\frac{\partial \widehat{f}_2^{(1)}}{\partial \rho} + \frac{\partial \widehat{f}_1^{(1)}}{\partial z} + \frac{\eta + a + 1}{\rho} \widehat{f}_2^{(1)} + \alpha A \widehat{f}_2^{(1)} + (2/\varsigma + \alpha C_e + \alpha \overline{\varphi}) \widehat{f}_3^{(1)} - \alpha v_0 \widehat{f}_3^{(2)} - \alpha v \widehat{f}_2^{(2)} = 0, \tag{3.23}$$

$$\frac{\partial \widehat{f}_1^{(1)}}{\partial \rho} - \frac{\partial \widehat{f}_2^{(1)}}{\partial z} + \frac{\eta - a}{\rho} \widehat{f}_1^{(1)} - \alpha A \widehat{f}_1^{(1)} + (2/\varsigma + \alpha C_e + \alpha \overline{\varphi}) \widehat{f}_4^{(1)} - \alpha v_0 \widehat{f}_4^{(2)} + \alpha v \widehat{f}_1^{(2)} = 0. \tag{3.24}$$

We cannot eliminate the terms with $1/\rho$ in all the equations but we can eliminate these terms in two of them. There are two possibilities. First, we can put $\eta = a$ obtaining equations akin in structure to Eqs. (4.6)–(4.9) of [2]. Since the number $\eta$ is a positive integer, zero or a positive half-integer, such must the constant $a$. By analogy with Eqs. (4.24)–(4.25) of [2] we make the replacement

$$\widehat{f}_1^{(1)} = g_1^{(1)}(\rho^2, z^2), \ \widehat{f}_2^{(1)} = \rho z g_2^{(1)}(\rho^2, z^2), \ \widehat{f}_3^{(1)} = z g_3^{(1)}(\rho^2, z^2), \ \widehat{f}_4^{(1)} = \rho g_4^{(1)}(\rho^2, z^2)$$

$$\widehat{f}_1^{(2)} = z g_1^{(2)}(\rho^2, z^2), \ \widehat{f}_2^{(2)} = \rho g_2^{(2)}(\rho^2, z^2), \ \widehat{f}_3^{(2)} = g_3^{(2)}(\rho^2, z^2), \ \widehat{f}_4^{(2)} = \rho z g_4^{(2)}(\rho^2, z^2),$$

$$\varphi = \varphi(\rho^2, z^2), \ A = \rho \gamma(\rho^2, z^2), \ v_0 = z u_0(\rho^2, z^2), \ v = \rho z u(\rho^2, z^2). \tag{3.25}$$

Now Eqs. (3.21)–(3.24) reduce to

$$\rho \frac{\partial g_4^{(1)}}{\partial \rho} + z \frac{\partial g_3^{(1)}}{\partial z} + 2(a + 1) g_4^{(1)} + g_3^{(1)} + \alpha \rho^2 \gamma g_4^{(1)} - \alpha(C_e + \overline{\varphi}) g_1^{(1)} + \alpha z^2 u_0 g_1^{(1)} - \alpha \rho^2 z^2 u g_4^{(2)} = 0,$$

$$\tag{3.26}$$

$$\frac{\partial g_3^{(1)}}{\rho \partial \rho} - \frac{\partial g_4^{(1)}}{z \partial z} - \alpha \gamma g_3^{(1)} - \alpha \left( C_e + \overline{\varphi} \right) g_2^{(1)} + \alpha u_0 g_2^{(2)} + \alpha u g_3^{(2)} = 0, \tag{3.27}$$

$$\rho \frac{\partial g_2^{(1)}}{\partial \rho} + \frac{\partial g_1^{(1)}}{z \partial z} + 2(a+1) g_2^{(1)} + \alpha \rho^2 \gamma g_2^{(1)} + \left( 2/\varsigma + \alpha C_e + \alpha \overline{\varphi} \right) g_3^{(1)} - \alpha u_0 g_3^{(2)} - \alpha \rho^2 u g_2^{(2)} = 0,$$
$$\tag{3.28}$$

$$\frac{\partial g_1^{(1)}}{\rho \partial \rho} - z \frac{\partial g_2^{(1)}}{\partial z} - g_2^{(1)} - \alpha \gamma g_1^{(1)} + \left( 2/\varsigma + \alpha C_e + \alpha \overline{\varphi} \right) g_4^{(1)} - \alpha z^2 u_0 g_4^{(2)} + \alpha z^2 u g_1^{(2)} = 0. \tag{3.29}$$

In the present case Eq. (3.11) acquires the form

$$\rho \frac{\partial g_4^{(2)}}{\partial \rho} + \frac{\partial g_3^{(2)}}{z \partial z} + 2(a+1) g_4^{(2)} + \alpha \rho^2 \gamma g_4^{(2)} - \alpha \left( C_e + \overline{\varphi} \right) g_1^{(2)} - \alpha u_0 g_1^{(1)} + \alpha \rho^2 u g_4^{(1)} = 0, \tag{3.30}$$

$$\frac{\partial g_3^{(2)}}{\rho \partial \rho} - z \frac{\partial g_4^{(2)}}{\partial z} - g_4^{(2)} - \alpha \gamma g_3^{(2)} - \alpha \left( C_e + \overline{\varphi} \right) g_2^{(2)} - \alpha z^2 u_0 g_2^{(1)} - \alpha z^2 u g_3^{(1)} = 0, \tag{3.31}$$

$$\rho \frac{\partial g_2^{(2)}}{\partial \rho} + z \frac{\partial g_1^{(2)}}{\partial z} + g_1^{(2)} + 2(a+1) g_2^{(2)} + \alpha \rho^2 \gamma g_2^{(2)} + \left( 2/\varsigma + \alpha C_e + \alpha \overline{\varphi} \right) g_3^{(2)}$$
$$+ \alpha z^2 u_0 g_3^{(1)} + \alpha \rho^2 z^2 u g_2^{(1)} = 0, \tag{3.32}$$

$$\frac{\partial g_1^{(2)}}{\rho \partial \rho} - \frac{\partial g_2^{(2)}}{z \partial z} - \alpha \gamma g_1^{(2)} + \left( 2 + \alpha C_e + \alpha \overline{\varphi} \right) g_4^{(2)} + \alpha u_0 g_4^{(1)} - \alpha u g_1^{(1)} = 0. \tag{3.33}$$

Further treatment of these equations can be carried out in much the same way as the treatment of Eqs. (4.28)− (4.35) of Ref. [2]. If $a = 0$, the last equations will differ from Eqs. (3.26)−(3.33) only in sign in front of $\gamma$, $\overline{\varphi}$ and $C_e$. In the event that Eqs. (3.26)–(3.29) are transformed to the type of Eqs. (4.52) and (A.1) of [2], one will have

$$\nu^2 = -\alpha C_e \left( 2/\varsigma + \alpha C_e \right). \tag{3.34}$$

The quantity $\nu^2$ will be positive if $-2/\zeta < \alpha C_e < 0$.

The second possibility concerning Eqs. (3.21)–(3.24) is to put $\eta = -a-1$. Since the number $\eta$ is a positive integer, zero or a positive half-integer, the constant $a$ must be a negative integer or half-integer satisfying the inequality $a \leq -1$. Instead of (3.25) we now make the replacement

$$\widehat{f}_1^{(1)} = \rho z g_1^{(1)}(\rho^2, z^2), \, \widehat{f}_2^{(1)} = g_2^{(1)}(\rho^2, z^2), \, \widehat{f}_3^{(1)} = \rho g_3^{(1)}(\rho^2, z^2), \, \widehat{f}_4^{(1)} = z g_4^{(1)}(\rho^2, z^2)$$

$$\widehat{f}_1^{(2)} = \rho g_1^{(2)}(\rho^2, z^2), \, \widehat{f}_2^{(2)} = z g_2^{(2)}(\rho^2, z^2), \, \widehat{f}_3^{(2)} = \rho z g_3^{(2)}(\rho^2, z^2), \, \widehat{f}_4^{(2)} = g_4^{(2)}(\rho^2, z^2),$$

$$\varphi = \varphi(\rho^2, z^2), \, A = \rho \gamma(\rho^2, z^2), \, v_0 = z u_0(\rho^2, z^2), v = \rho z u(\rho^2, z^2). \tag{3.35}$$

Equations (3.21)–(3.24) and (3.11) now yield

$$\frac{\partial g_4^{(1)}}{\rho \partial \rho} + \frac{\partial g_4^{(1)}}{z \partial z} + \alpha \gamma g_4^{(1)} - \alpha \left( C_e + \overline{\varphi} \right) g_1^{(1)} + \alpha u_0 g_1^{(2)} - \alpha u g_4^{(2)} = 0, \tag{3.36}$$

$$\rho \frac{\partial g_3^{(1)}}{\partial \rho} - z \frac{\partial g_4^{(1)}}{\partial z} - 2a g_3^{(1)} - g_4^{(1)} - \alpha \rho^2 \gamma g_3^{(1)} - \alpha (C_e + \overline{\varphi}) g_2^{(1)} + \alpha z^2 u_0 g_2^{(2)} + \alpha \rho^2 z^2 u g_3^{(2)} = 0, \quad (3.37)$$

$$\frac{\partial g_2^{(1)}}{\rho \partial \rho} + z \frac{\partial g_1^{(1)}}{\partial z} + g_1^{(1)} + \alpha \gamma g_2^{(1)} + (2/\varsigma + \alpha C_e + \alpha \overline{\varphi}) g_3^{(1)} - \alpha z^2 u_0 g_3^{(2)} - \alpha z^2 u g_2^{(2)} = 0. \quad (3.38)$$

$$\rho \frac{\partial g_1^{(1)}}{\partial \rho} - \frac{\partial g_2^{(1)}}{z \partial z} - 2a g_1^{(1)} - \alpha \rho^2 \gamma g_1^{(1)} + (2/\varsigma + \alpha C_e + \alpha \overline{\varphi}) g_4^{(1)} - \alpha u_0 g_4^{(2)} + \alpha \rho^2 u g_1^{(2)} = 0, \quad (3.39)$$

$$\frac{\partial g_4^{(2)}}{\rho \partial \rho} + z \frac{\partial g_3^{(2)}}{\partial z} + g_3^{(2)} + \alpha \gamma g_4^{(2)} - \alpha (C_e + \overline{\varphi}) g_1^{(2)} - \alpha z^2 u_0 g_1^{(1)} + \alpha z^2 u g_4^{(1)} = 0, \quad (3.40)$$

$$\rho \frac{\partial g_3^{(2)}}{\partial \rho} - \frac{\partial g_4^{(2)}}{z \partial z} - 2a g_3^{(2)} - \alpha \rho^2 \gamma g_3^{(2)} - \alpha (C_e + \overline{\varphi}) g_2^{(2)} - \alpha u_0 g_2^{(1)} - \alpha \rho^2 u g_3^{(1)} = 0, \quad (3.41)$$

$$\frac{\partial g_2^{(2)}}{\rho \partial \rho} + \frac{\partial g_1^{(2)}}{z \partial z} + \alpha \gamma g_2^{(2)} + (2/\varsigma + \alpha C_e + \alpha \overline{\varphi}) g_3^{(2)} + \alpha u_0 g_3^{(1)} + \alpha u g_2^{(1)} = 0, \quad (3.42)$$

$$\rho \frac{\partial g_1^{(2)}}{\partial \rho} - z \frac{\partial g_2^{(2)}}{\partial z} - 2a g_1^{(2)} - g_2^{(2)} - \alpha \rho^2 \gamma g_1^{(2)} + (2/\varsigma + \alpha C_e + \alpha \overline{\varphi}) g_4^{(2)}$$
$$+ \alpha z^2 u_0 g_4^{(1)} - \alpha \rho^2 z^2 u g_1^{(1)} = 0. \quad (3.43)$$

These equations are similar in structure to Eqs. (3.26)–(3.33) and can be treated analogously although the roles of the functions $g_i^{(1,2)}$ are different as compared to Eqs. (3.26)–(3.33).

Summarizing the analysis of Eqs. (3.21)–(3.24) we see that the constant $a$ in (3.15) must be an integer, zero or a half-integer subject to the restrictions $a \geq 0$ or $a \leq -1$. Only the interval $-1 < a < 0$ is excluded. In actual fact, the value $a = -1/2$ alone is prohibited.

We now turn to Eqs. (2.21)–(2.22). The terms in the equations that contain derivatives are fully analogous to the relevant terms in (3.10)–(3.11). We make the substitution of (3.15) that leads to equations of the type (3.16)–(3.19) with $b$ instead of $a$. Next, we make the replacement akin to (3.20):

$$F_i^{(1,2)} = \rho^{\eta_p} \widehat{F}_i^{(1,2)}, \quad (3.44)$$

and obtain equations of the type (3.21)–(3.24). We shall again have two possibilities: $\eta_p = b$ and $\eta_p = -b-1$. The first possibility leads to equations similar to (2.48)–(2.55), the second possibility to analogous equations as in (3.36)–(3.43). As a result, the constant $b$ in (3.15) as well as the constant $a$ must be an integer, zero or a half-integer, excluding the value $b = -1/2$.

*3.2. Spins of nuclei*

In this paper we restrict our consideration to the electronic field concentrated completely inside nuclei, that is to say, we shall consider solely "naked" nuclei devoid of the external electronic envelope. The angular momentum **M** of a nucleus, the nuclear spin, is given by (3.8). Similarly to [2] and Sec. 2 above it can be shown that the first two integrals in (3.8) vanish for the stationary and axially symmetric solutions under consideration if Eq. (3.12) is taken into account. We introduce (3.15) into the last two integrals and integrate over the angle $\dot{\phi}$ to see that there is only one nonzero component $M_z$ of the vector **M**. With use made of (3.6) the component in the ordinary units becomes

$$M_z = \left(a + \frac{1}{2}\right)\hbar N_e + \left(b + \frac{1}{2}\right)\hbar N_p \,. \tag{3.44}$$

The spin is closely related to the magnetic moment **μ**. Analogously with Eq. (2.43) we find that **μ** has only one component

$$\mu_z = \int\limits_{(\infty)} \rho\Big[-f_1^{(1)}f_4^{(1)} + f_2^{(1)}f_3^{(1)} + f_1^{(2)}f_4^{(2)} - f_2^{(2)}f_3^{(2)}$$

$$+ F_1^{(1)}F_4^{(1)} - F_2^{(1)}F_3^{(1)} - F_1^{(2)}F_4^{(2)} + F_2^{(2)}F_3^{(2)} - \frac{\kappa}{2}\left(\frac{\partial\Delta_\rho}{\partial z} - \frac{\partial\Delta_z}{\partial\rho}\right)\Big]dV \,. \tag{3.45}$$

It is worth remarking that the vector **Δ** of (2.19) does not depend on the factor $\exp(ib\dot{\phi})$ in (3.15). Equations (3.44) and (3.45) demonstrate that the relation between $\mu_z$ and $M_z$ is not so simple. This is exampled by the proton for which $\mu_z \approx 2.793\ \mu_N$ whereas $M_z = \hbar/2$.

As established above the constants $a$ and $b$ in (3.44) must be integers, zero or half-integers excluding the values $a = -1/2$ and $b = -1/2$. Now one can understand this exclusion. If $a = -1/2$ in (3.44), the spin of the electronic field, the case of one electron inclusive when $N_e = 1$, will be nil. In this instance there will be no rotational motion in the electron in order to compensate the repulsive Coulomb forces inside the electron according to Ref. [2] so that the electron will not exist. If $b = -1/2$, the same concerns the proton. It is interesting to note that the electron spin [$N_e = 1$ and $N_p = 0$ in (3.44)] is not necessarily equal to $\hbar/2$. It can be equal even to an integer ($\times\hbar$). Of course, this will correspond to excited states of the electron if $N_e = 1$ and $N_p = 0$.

The number of protons $N_p$ in a nuclide $^A_Z X$ is equal to the mass number $A$, and the number of electrons $N_e$ is $A - Z$ (previously this was the number of neutrons). According to (3.44) the spin of a nucleus can be different depending on the numbers $a$ and $b$. We have so selected these numbers that the resulting spin would coincide with the experimental value taken from [7] or, if it is impossible, would be as close to the experimental value as possible. The results are presented in Table 1 for different stable nuclides taken as examples (the neutron is also included although it is unstable). Two values of $a$ and $b$ for each nuclide are such that they give one and the same $\eta$ in (3.20) and one and the same $\eta_p$ in (3.44) so that they are relevant to different orientations of the nucleus. It should be remarked that one and the same $M_z$ can be obtained with different $a$ and $b$. The table contains the least possible values of $|a|$ and $|b|$ although sometimes close values are possible. For example, in the case of the deuteron $^2$H the value $M_z/\hbar = 1$ is also produced by $a = -3/2$ and $b = 1/2$ and $M_z/\hbar = -1$ by $a = 1/2$ and $b = -3/2$. In the table are placed the values that yield $\eta_p = 0$ for $^2$H. The numbers $a$ and $b$ are to be found when solving the above equations.

### Table 1. Spins of nuclei

| Nucleus | $a$ | $b$ | $M_z/\hbar$ | $M_z/\hbar$, Ref. [7] | Nucleus | $a$ | $b$ | $M_z/\hbar$ | $M_z/\hbar$, Ref. [7] |
|---|---|---|---|---|---|---|---|---|---|
| Neutron | 1/2 | −1 | 1/2 | 1/2 | $^{26}$Mg | 6 | −4 | 0 | 0 |
|  | −3/2 | 0 | −1/2 |  |  | −7 | 3 | 0 |  |
| $^2$H | 3/2 | −1 | 1 | 1 | $^{64}$Zn | 31/2 | −9 | 0 | 0 |
|  | −5/2 | 0 | −1 |  |  | −33/2 | 8 | 0 |  |
| $^6$Li | 1/2 | −1 | 0 | 1 | $^{107}$Ag | −21 | 11 | 1/2 | 1/2 |
|  | −3/2 | 0 | 0 |  |  | 20 | −12 | −1/2 |  |
| $^7$Li | 5/2 | −2 | 3/2 | 3/2 | $^{136}$Ba | 8 | −11/2 | 0 | 0 |
|  | −7/2 | 1 | −3/2 |  |  | −9 | 9/2 | 0 |  |
| $^{10}$B | 1/2 | −1 | 0 | 3 | $^{182}$W | 45 | −55/2 | 0 | 0 |
|  | −3/2 | 0 | 0 |  |  | −46 | 53/2 | 0 |  |
| $^{11}$B | 5/2 | −2 | 3/2 | 3/2 | $^{183}$W | −24 | 27/2 | 1/2 | 1/2 |
|  | −7/2 | 1 | −3/2 |  |  | 23 | −29/2 | −1/2 |  |
| $^{14}$N | 1/2 | −1 | 0 | 1 | $^{197}$Au | −8 | 4 | 3/2 | 3/2 |
|  | −3/2 | 0 | 0 |  |  | 7 | −5 | −3/2 |  |
| $^{23}$Na | 5/2 | −2 | 3/2 | 3/2 | $^{206}$Pb | 51 | −63/2 | 0 | 0 |
|  | −7/2 | 1 | −3/2 |  |  | −52 | 61/2 | 0 |  |
| $^{24}$Mg | 1/2 | −1 | 0 | 0 | $^{207}$Pb | 26 | −33/2 | 1/2 | 1/2 |
|  | −3/2 | 0 | 0 |  |  | −27 | 31/2 | −1/2 |  |
| $^{25}$Mg | 9/2 | −3 | 5/2 | 5/2 | $^{208}$Pb | 103/2 | −32 | 0 | 0 |
|  | −11/2 | 2 | −5/2 |  |  | −105/2 | 31 | 0 |  |

Inspecting the table we see that the spin given by (3.44) and the experimental one do not coincide only for the nuclides $^6_3\text{Li}$, $^{10}_5\text{B}$ and $^{14}_7\text{N}$. Interestingly enough, in all of these nuclides (and only in them if stable nuclides alone are taken into account) the mass number $A$ is even while the atomic number $Z$ is odd. The discrepancy between the theory and experiment may be explained as follows. Although the spin of the nuclei is nil according to (3.44), their magnetic moment can be distinct from zero according to (3.45). The spin is measured by investigating motion of particles in magnetic fields so that it is the magnetic moment that plays a leading role in the motion. The existence of the magnetic moment can imitate a nonzero spin. Besides, account must be taken of the fact that the spin of the "naked" nuclei considered may differ from the spin of the nuclei surrounded by an electronic envelope because this envelope affects the electronic field inside the nuclei as mentioned above, whereas it is unclear as to whether the tables of [7] contain the spin of "naked" nuclei or the spin of nuclei with their electronic envelopes. It should be added that we imply only axially symmetric solutions to the equations whereas other solutions are possible as well. In connection with Table 1 it is worthy of remark also that, when the values of $|a|$ and $|b|$ are considerable, the numbers $\eta$ in (3.20) and $\eta_p$ in (3.44) will be large as well. In this instance the density of matter near the axis $\rho = 0$ is extremely small and the nuclei have the shape of a torus.

### 3.3. The constant $\zeta$

The dimensionless equations of (3.10)–(3.13) and (2.21)–(2.22) contain three dimensionless constants $\alpha$, $\kappa$ and $\zeta$. The fine-structure constant $\alpha$ is determined from the condition of the existence of the electron according to Ref. [2]. The constant $\kappa$ is determined from the condition of the existence of the proton according to Sec. 2 above. The third constant $\zeta$ appears when the electronic and protonic fields are present simultaneously. The simplest system in which both the fields are present is a hydrogen atom. However a spatially bounded solution to the Dirac equation relevant to the hydrogen atom exists even if the electronic self-field is disregarded and the proton is considered to be a point. Corrections to the solution due to the electronic self-field and the structure of the proton can be sought by applying a perturbation theory to the above equations (cf. Ref. [3]). The constant $\zeta$ in these calculations will be arbitrary. The next simple system is a neutron. The neutron is, however, unstable whereas we imply only stationary and therefore stable systems.

We now turn to a deuteron that consists of two protons and one electron. The stability of the deuteron proves that two protons are capable of holding an electron inside the nucleus while one proton is not able to do so, which is demonstrated by the instability of the neutron. We put $N_e = 1$ and $N_p = 2$ in (3.6) and look for a particle-like and physically admissible solution to the above equations, the solution relevant to the deuteron. The solution should exist only if the constant $\zeta$ has a well-defined value. Of course the solution and the value of $\zeta$ can be found only numerically. As mentioned above, $\zeta \approx 1840$.

If the constant $\zeta$ is found when considering the deuteron, one may expect that the same $\zeta$ will produce other stationary solutions with other values of $N_e$ and $N_p$, the solutions that will be relevant to other stable nuclei. It will be recalled that we have other constants at our disposition, namely, the constants $C_e$ and $C_p$ and the numbers $a$ and $b$. It is conceivable that one will succeed in proving that the stationary solutions exist only if the numbers $N_e$ and $N_p$ lie in definite ranges. This will determine the region of stable nuclides.

## 4. Mesons

In what follows we shall be in need of an antiproton. The antiproton $\bar{p}$ can be described by analogy with the positron considered in Ref. [2]. If one introduces bispinors $\Psi'_1$ and $\Psi'_2$ according to $\Psi_1 = C\overline{\Psi}'_2$ and $\Psi_2 = C\overline{\Psi}'_1$, the bispinors $\Psi'_1$ and $\Psi'_2$ will obey the same equations of (2.4)–(2.8) and (2.12) with, however, another sign in front of $A_\mu$. Thus, if the system of equations of Sec. 2 has a solution $\Psi_1$, $\Psi_2$, $A_\mu$, $V_\mu$, the system has the solution $C\overline{\Psi}_2$, $C\overline{\Psi}_1$, $-A_\mu$, $V_\mu$ as well. The signs in front of $A_\mu$ demonstrate that the solutions are relevant to particles with opposite signs of the charge, that is, to protons and antiprotons respectively. The sign in the normalization of (2.11) also changes for the antiprotons as for the positron [2]. As a result, the equations of (3.6) are to be rewritten now as

$$\int\limits_{(\infty)} (\psi_1^* \psi_1 - \psi_2^* \psi_2)dV = N_e - N_{\bar{e}}, \quad \int\limits_{(\infty)} (\Psi_1^* \Psi_1 - \Psi_2^* \Psi_2)dV = N_p - N_{\bar{p}}, \tag{4.1}$$

where $N_{\bar{e}}$ and $N_{\bar{p}}$ are the number of positrons and antiprotons respectively in the system. These equations embody the conservation of the leptonic and baryonic charges. The numbers $N_e$ and $N_{\bar{e}}$ include also the number of other (heavy) leptons and antileptons because they can be regarded as excited states of the electron or positron according to [2]. At the same time the number of neutrinos and antineutrinos does not figure in (4.1) for they are only a "building waste» remaining after processes with the other leptons [3]. In conformity with the end of Sec. 2.1 the numbers $N_p$ and $N_{\bar{p}}$ include also the number of other baryons and antibaryons present in the system. Consequently, Eq. (4.1) can be recast in a more general form as

$$\int\limits_{(\infty)} (\psi_1^*\psi_1 - \psi_2^*\psi_2) dV = N_1 - N_{\bar{1}}, \quad \int\limits_{(\infty)} (\Psi_1^*\Psi_1 - \Psi_2^*\Psi_2) dV = N_b - N_{\bar{b}}, \qquad (4.2)$$

where $N_l$, $N_{\bar{l}}$, $N_b$ and $N_{\bar{b}}$ are respectively the number of leptons (without neutrinos and antineutrinos), antileptons, baryons and antibaryons in the system.

The annihilation of an electron and a positron was considered in Ref. [2]. From the point of view of (4.1) we have in this example that $\Psi_1 = \Psi_2 = N_p = N_{\bar{p}} = 0$ and $N_e = N_{\bar{e}} = 1$. Now the first equation in (4.1) becomes

$$\int\limits_{(\infty)} (\psi_1^*\psi_1 - \psi_2^*\psi_2) dV = 0. \qquad (4.3)$$

If we have $\psi_1 \neq 0$, $\psi_2 \neq 0$ first in (4.3), after the annihilation we have $\psi_1 = \psi_2 = 0$. Consequently, the electronic field disappears being converted into an electromagnetic wave described by Eq. (3.8) of [2] with $\psi_1 = \psi_2 = 0$. Equation (4.3) admits another interpretation. If we have first an electromagnetic wave alone ($\psi_1 = \psi_2 = 0$), it is able to create an electron-positron pair where $\psi_1 \neq 0$ and $\psi_2 \neq 0$. This does not contradict Eq. (4.3) or (4.1) with $N_e = N_{\bar{e}} = 1$. The creation is possible when the frequency of the wave is $\omega = 2mc^2/\hbar$. This process is due to a resonance and is discussed in Sec. 7 of [8]. In a similar way, a protonic field can be generated. Thus, the electronic and protonic fields can come into being even if they were absent initially with the proviso that (4.2) is fulfilled.

Mesons do not exist in nature on their own. They appear only as short-lived products of collisions between divers particles. Let us consider an example:

$$p + \bar{p} \to \pi^+ + \pi^-, \qquad (4.4)$$

where, for simplicity's sake, we have taken only two pions. To understand the nature of the pions appearing in (4.4) we write down their possible decays:

$$\pi^+ \to \mu^+ + \nu_\mu, \quad \pi^- \to \mu^- + \bar{\nu}_\mu. \qquad (4.5)$$

From these we see that the pions must contain an electronic field because the muons contain it while the first equation of (4.2) with the right-hand side different from zero holds for the processes given by (4.5). At the same time the pions should contain something else as long as they are not leptons. We see from (4.4) that the pions should also incorporate a protonic field that comes from the proton and antiproton. As a result, we have $N_l = 1$, $N_{\bar{l}} = N_b = N_{\bar{b}} = 0$ in (4.2) for the negative pion $\pi^-$ and $N_{\bar{l}} = 1$, $N_l = N_b = N_{\bar{b}} = 0$ for the positive pion $\pi^+$ while $\Psi_1 \neq 0$ and $\Psi_2 \neq 0$ for both of them. It is worthy of remark that we have the same numbers for a lepton or an antilepton but $\Psi_1 = \Psi_2 = 0$ for them. In the decays of (4.5) the protonic field characterized by $\Psi_{1,2}$ disappears.

Other mesons eventually transform into pions and muons. Therefore they are also composed of the electronic and protonic fields. The mesons can be regarded as short-lived splinters resulting from collisions of various particles.

## 5. Concluding remarks

From the results of this paper it follows that elementary particles interact only via an electromagnetic field. Of course there is a gravitational interaction too, but it is believed that the last interaction plays an insignificant role for the elementary particles. The equations of Sec. 3 that describe nuclei contain only interaction terms due to an electromagnetic potential $A_\mu$. The electromagnetic interaction in a nucleus is so modified that it imitates the presence of another interaction, the strong interaction. The relative strength of the strong interaction with respect to the electromagnetic one is estimated to be $10^2$–$10^3$. The Lagrangian for the proton of (2.3) contains derivatives of $A_\mu$ whereas they are absent in the electron Lagrangian. The protonic field is concentrated in a region of size of order $\lambdabar_p$ and thereby the spatial derivatives are of the order $A_\mu/\lambdabar_p$ while the same derivatives in the electronic field are of the order $A_\mu/\lambdabar$. Therefore the ratio of these derivatives is of the order $\lambdabar/\lambdabar_p \approx m_p/m \sim 10^3$. We see that the derivatives of $A_\mu$ present in the Lagrangian of (2.3) can explain the strength ascribed to the strong interaction. The weak interaction is also absent in nature. To explain the beta decay of a neutron ($n \rightarrow p + e^- + \bar{\nu}_e$) it is unnecessary to invoke the weak interaction. The neutron is composed of a proton and an electron. This system is unstable and disintegrates into its constituents. The appearance of the (anti)neutrino in the process as a "building waste» is explained in Sec. 5 of [3]. Other processes that can be explicated without resorting to the idea of the weak interaction are given by (4.5) where the protonic field $\Psi_{1,2}$ disappears in the course of the processes and nothing else, which is completely described by the equations of Sec. 3.

In nature there exist only two stable charged elementary particles, the electron and proton (with their antiparticles). This suggests that any form of matter can be represented in terms of electronic and protonic fields. In particular, metaphorically speaking nuclei are composed of electrons and protons although there are no separate electrons and no separate protons in the nuclei but a unified electronic field and unified protonic field are present in them. Though the hypothesis of the electron-proton constitution of the nuclei was the first hypothesis concerning the structure of the nuclei, it was rejected because of a naïve consideration of the spins. We have demonstrated in Sec. 3 that the spin of a proton or an electron is not obligatory equal to $\hbar/2$.

Their spin in units of $\hbar$ can be any half-integer or integer excluding zero. This enables one to explain the experimentally observed spin of the nuclei from the proposed point of view.

In this paper we restricted our consideration to the electronic field concentrated completely inside nuclei, that is to say, we considered solely "naked" nuclei devoid of the external electronic envelope. In the case of a neutral atom the electronic field exists in the nucleus and in the electronic envelope of the atom and the field is unique. This fact essentially complicates consideration of the electronic envelope in the present approach, which cannot be done without considering the nucleus. In Ref. [3] it was suggested that the electronic envelope contains clots resembling separate electrons. If this is the case, the electronic field cannot be independent of the angle $\dot{\varphi}$ in cylindrical coordinates, which amounts to saying that the field will no longer be axially symmetric. Hence, if one wants to study the electronic envelope of the atom in the framework of the present approach, one should look for $\dot{\varphi}$-dependent solutions to the equations obtained at the outset of Sec. 3. Of course, approximate methods are possible in this question in line with Ref. [3], the methods that do not take the structure of the nucleus into account.

The equations of this paper contain three dimensionless constants $\alpha$, $\kappa$ and $\zeta$. The value of the fine-structure constant $\alpha$ is found from the condition of the existence of the electron according to Ref. [2]. The constant $\kappa$ is determined from the condition of the existence of the proton, which is considered in Sec. 2 of this paper. The third constant $\zeta$ that gives the relation between the masses of the electron and proton follows from the stability of the deuteron and other stable nuclei. We see that the present theory of elementary particles takes no constants from experiment, but, on the contrary, the theory explains the numerical value of the occurring constants.

According to Ref. [8] there are no photons in nature. Reference [2] demonstrates that the leptons are described by an electronic field. The ground state of the field corresponds to an electron while excited states to other leptons which are unstable for this reason. The present paper shows that the baryons are relevant to a protonic field. The ground state of the last field corresponds to a proton while excited states to other baryons. The mesons are composed of electronic and protonic fields. They are short-lived splinters resulting from collisions of various particles. Motion, collisions and scattering of the particles can be investigated with the help of equations of Sec. 3.

It follows from Ref. [2] that the neutrinos are described by the electronic field and correspond to objects moving at the speed of light. The question arises as to whether there are similar objects moving at the speed of light and relevant to the protonic field. In the simplest case such proton neutrinos will be described by equations of Sec. 2 if one assumes that the dependence of all functions upon coordinates and time is of the form $F = F(x, y, z - ct)$ {cf. Eqs. (5.3)–(5.4) of [2]}. Then Eqs. (2.21)–(2.24) will be of the form

$$-i\frac{\partial \tilde{\Psi}_1}{\partial \tilde{z}} + i\boldsymbol{\alpha}\tilde{\nabla}\tilde{\Psi}_1 + \alpha\tilde{\mathbf{A}}\boldsymbol{\alpha}\tilde{\Psi}_1 - \alpha\tilde{\varphi}\tilde{\Psi}_1 - \beta\tilde{\Psi}_1 - i\alpha\kappa\left(\tilde{\mathbf{E}}\boldsymbol{\gamma} + \tilde{\mathbf{H}}\boldsymbol{\beta}\right)\tilde{\Psi}_1 - \alpha\tilde{V}_0\tilde{\Psi}_2 + \alpha\tilde{\mathbf{V}}\boldsymbol{\alpha}\tilde{\Psi}_2 = 0, \quad (5.1)$$

$$-i\frac{\partial \tilde{\Psi}_2}{\partial \tilde{z}} + i\boldsymbol{\alpha}\tilde{\nabla}\tilde{\Psi}_2 + \alpha\tilde{\mathbf{A}}\boldsymbol{\alpha}\tilde{\Psi}_2 - \alpha\tilde{\varphi}\tilde{\Psi}_2 - \beta\tilde{\Psi}_2 - i\alpha\kappa\left(\tilde{\mathbf{E}}\boldsymbol{\gamma} + \tilde{\mathbf{H}}\boldsymbol{\beta}\right)\tilde{\Psi}_2 + \alpha\tilde{V}_0\tilde{\Psi}_1 - \alpha\tilde{\mathbf{V}}\boldsymbol{\alpha}\tilde{\Psi}_1 = 0, \quad (5.2)$$

$$\nabla^2\tilde{\varphi} - \frac{\partial}{\partial \tilde{z}}\operatorname{div}\tilde{\mathbf{A}} + 4\pi\left(\tilde{\Psi}_1^*\tilde{\Psi}_1 - \tilde{\Psi}_2^*\tilde{\Psi}_2\right) + 4\pi\kappa\operatorname{div}\tilde{\boldsymbol{\Gamma}} = 0, \quad (5.3)$$

$$\nabla^2\tilde{\mathbf{A}} - \frac{\partial^2\tilde{\mathbf{A}}}{\partial \tilde{z}^2} - \nabla\operatorname{div}\tilde{\mathbf{A}} + \nabla\frac{\partial\tilde{\varphi}}{\partial \tilde{z}} + 4\pi\left(\tilde{\Psi}_1^*\boldsymbol{\alpha}\tilde{\Psi}_1 - \tilde{\Psi}_2^*\boldsymbol{\alpha}\tilde{\Psi}_2\right) + 4\pi\kappa\frac{\partial\tilde{\boldsymbol{\Gamma}}}{\partial \tilde{z}} - 4\pi\kappa\operatorname{curl}\tilde{\boldsymbol{\Delta}} = 0. \quad (5.4)$$

Further analysis is required to find out whether these equations admit physically reasonable solutions, in which case one will have the proton neutrinos with their properties. Solutions to the equations of Sec. 2 can be sought in a more general form $F = F(x, y, z - ct, t)$ taking account of evolution of the neutrinos in flight. Separate investigation is needed to elucidate how the proton neutrinos can be observed experimentally if they exist.